\documentclass[final,5p,times,twocolumn,number]{elsarticle}
\usepackage{amsmath,booktabs,graphicx,dblfloatfix,capt-of,microtype}
\usepackage[hidelinks]{hyperref}
\usepackage{xcolor}
\definecolor{revisionblue}{RGB}{0,65,165}
\definecolor{noveltygreen}{RGB}{0,128,0}
\newif\ifmarkchanges\markchangesfalse

\definecolor{latestmagenta}{RGB}{0,0,0}

\journal{Nuclear Physics B}

\date{October 21, 2026}
\begin{document}
\begin{frontmatter}
\title{Color-Reconnection Effects on the \texorpdfstring{$n_{\rm MPI}$--$N_{\rm ch}$}{nMPI--Nch} Conditional Multiplicity Fluctuations in PYTHIA 8 pp Collisions}
\author[avc]{Subir Sarkar\corref{cor1}}\ead{subir.sarkar@avcollege.ac.in}
\cortext[cor1]{Corresponding author}
\affiliation[avc]{organization={Department of Physics, Arya Vidyapeeth College (Autonomous)},city={Guwahati},postcode={781016},state={Assam},country={India}}
\author[bu]{Kalyan Dey}
\affiliation[bu]{organization={Department of Physics, Bodoland University},city={Kokrajhar},state={Assam},country={India}}

\begin{abstract}

Color reconnection (CR) changes both the mean and the dispersion of charged-particle multiplicity in pp collisions. We study the conditional distribution $P(N_{\rm ch}\mid n_{\rm MPI}=m)$ at fixed PYTHIA interaction count for $\sqrt{s}=2.76$, 7, and 13~TeV, using charged particles in $|\eta|<0.8$ and $0.15<p_{\rm T}<20$~GeV/$c$. CR is switched on or off while the remaining tune parameters are held fixed. Enabling CR reduces the conditional mean by a larger fraction than the standard deviation. The absolute width therefore decreases, while the coefficient of variation increases in every retained matched MPI bin. This ordering also holds in independent 13~TeV samples with Monash, A14, and CMS CP5. Comparisons with ALICE mean-transverse-momentum data and simulated activity estimators relate these conditional fluctuations to experimentally accessible final-state observables.

\end{abstract}
\end{frontmatter}

\section{Introduction}
\label{sec:intro}

Multiparton interactions (MPI) allow several parton--parton scatterings in a single pp collision. They help describe underlying-event activity and broad multiplicity distributions beyond a single hard scattering~\cite{SjostrandVanZijl1987,Bierlich:2022pfr}. The final charged multiplicity is also shaped by showers, beam remnants, color connections, hadronization, and decays. Its relation to the MPI count therefore depends on the full event evolution.

Color reconnection rearranges the color connections between partons before string fragmentation. In the MPI-based model used here, softer interaction systems can reconnect to harder ones, reducing the string-length measure~\cite{Bierlich:2022pfr}. This modifies particle production without changing the count of already generated interactions. CR is known to affect multiplicity, mean transverse momentum, and identified-particle observables~\cite{ALICE:2013rdo,OrtizVelasquez2013,CMSCRTune}.

Throughout this work, $n_{\rm MPI}$ denotes PYTHIA's internal interaction count. This model-dependent quantity provides a reference for studying how showers, hadronization, and decays translate partonic activity into charged-particle production. Its interpretation depends on the generator version, tune, process selection, and MPI settings; it is not directly measurable. Supplementary Material S1 specifies the implementation.

The relation between MPI activity and final-state multiplicity, including its dependence on CR, has been studied in several contexts. Weber et al.~\cite{Weber:2018ddv} examine multiplicity production and fluctuations at a given MPI count in quarkonium studies. Prasad et al.~\cite{Prasad:2024gqq} compare activity classifiers with and without CR and with rope hadronization, while Prasad et al.~\cite{Prasad:2025yfj} review their MPI sensitivity and selection biases. These studies motivate a joint examination of the conditional mean, absolute width, and width relative to the mean.

Here we quantify how CR changes the mean and dispersion of charged multiplicity at fixed MPI count. We evaluate the conditional mean $\mu_m$, absolute standard deviation $\sigma_m$, and coefficient of variation $C_m=\sigma_m/\mu_m$ together, comparing CR ON and CR OFF at the same integer $m=n_{\rm MPI}$. This comparison distinguishes a change in absolute width from a change in width relative to the local mean. We examine whether the resulting ordering persists at three collision energies and in independent tune configurations at 13~TeV, and assess its sensitivity to the stated selections.

We first use inclusive distributions to motivate conditioning on MPI count, then compare the conditional means and widths. The ALICE comparison and final-state estimators provide supporting context, followed by checks of tune and selection dependence.

\section{Simulation, definitions, and scope}
\label{sec:simulation}

The baseline consists of five million events for each energy and CR state, with the nominal setup PYTHIA 8.316, \texttt{SoftQCD:all}, and Monash 2013~\cite{MonashTune}. The CR variation uses \texttt{ColourReconnection:reconnect = off}, with the remaining tune parameters held fixed in the intended setup. Historical production-time seed values and effective lifetime/decay settings were not retained for the baseline samples, which limits exact generator replay. S1 and S2 describe the available production records and reproducible analysis of the retained events.

\subsection{MPI and CR settings}

The independent 13~TeV tune comparison uses Monash (\texttt{Tune:pp = 14}), central A14 NNPDF2.3 LO (\texttt{Tune:pp = 21})~\cite{ATLASA14Tune}, and the published CMS CP5 parameters~\cite{CMSCP5Tune,CMSCP5SettingsCard}. Table~\ref{tab:settings} gives the key settings. Each tune and CR state has five million generated events.

\begin{table*}[!tbp]\centering\footnotesize
\caption{\label{tab:settings}Key settings for the independent 13~TeV tune comparison in PYTHIA 8.316. All use \texttt{SoftQCD:all = on}; only the CR physics switch changes within each tune pair. ISR and FSR denote initial- and final-state radiation; running orders 1 and 2 mean one- and two-loop running. Settings and provenance are detailed in Supplementary Materials S1 and S2.}
{\ifmarkchanges\color{revisionblue}\else\color{black}\fi\begin{tabular}{lccc}\toprule
Setting & Monash & A14 & CMS CP5\\\midrule
\texttt{Tune:pp} (starting point)&14&21&14 + overrides\\
PDF family &NNPDF2.3 LO QCD+QED&NNPDF2.3 LO QCD+QED&NNPDF3.1 NNLO\\
MPI $p_{\rm T0}^{\rm ref}$ [GeV]&2.28&2.09&1.41\\
MPI reference energy [GeV]&7000&7000&7000\\
MPI energy exponent&0.215&0.215&0.03344\\
\texttt{MultipartonInteractions:bProfile}&3&3&2\\
Overlap exponent (profile 3)&1.85&1.85&--\\
Core radius / fraction (profile 2)&--&--&0.7634 / 0.63\\
\texttt{ColourReconnection:mode}&0&0&0\\
\texttt{ColourReconnection:range}&1.80&1.71&5.176\\
$\alpha_s(M_Z)$: hard process / running order&0.130 / 1&0.140 / 1&0.118 / 2\\
$\alpha_s(M_Z)$: ISR / running order&0.1365 / 1&0.127 / 1&0.118 / 2\\
$\alpha_s(M_Z)$: FSR / running order&0.1365 / 1&0.127 / 1&0.118 / 2\\
$\alpha_s(M_Z)$: MPI / running order&0.130 / 1&0.126 / 1&0.118 / 2\\\bottomrule
\end{tabular}}
\end{table*}

The MPI regularization scale controls the low-$p_{\rm T}$ interaction rate, and its energy dependence controls how MPI activity changes with collision energy. The regularized form and energy-scaling equation are given in Supplementary Material S1, following Bierlich et al.~\cite{Bierlich:2022pfr}.

The matter profile controls the impact-parameter dependence of MPI activity. Monash and A14 use profile 3, specified by an overlap exponent; CP5 uses a double-Gaussian profile, specified by a core radius and fraction. All three use MPI-based CR mode 0, with a range parameter controlling the reconnection probability of softer systems. The profile forms and full settings are given in Supplementary Material S1.

CP5 uses the NNPDF3.1 NNLO central PDF through LHAPDF~\cite{LHAPDF6}, with two-loop running and $\alpha_s(M_Z)=0.118$ in all four coupling sectors. The full-precision CMS tune settings are implemented in PYTHIA 8.316, rather than the PYTHIA 8.226 environment of the original CMS production. Both CR states use the default PYTHIA decay prescription and independent recorded seeds. S1 summarizes the settings, and S2 supplies the available production records.

\subsection{Particle and event definitions}
\label{sec:selections}

The analysis multiplicity is

\begin{equation}
N_{\rm ch}=N_{\rm ch}(|\eta|<0.8,\;0.15<p_{\rm T}<20~{\rm GeV}/c).
\end{equation}

An INEL$>0$-like event contains at least one charged final-state particle in $|\eta|<1$. Inclusive distributions use this event selection without an $n_{\rm MPI}$ cut. The conditional response study additionally requires $n_{\rm MPI}>2$ and compares common integer bins containing at least 500 events in each CR state.

Charged final-state particles are selected from the generator records; implementation details are given in Supplementary Material S1. No detector simulation is applied. The ALICE comparison requires $N_{\rm ch}>0$ and no MPI cut, with the primary-particle limitation discussed in Sec.~\ref{sec:alice}. The estimator selections are specified in Secs.~\ref{sec:estimators} and \ref{sec:tunes}.

\subsection{AI-assisted code development and validation}
\label{sec:ai_methods}
OpenAI Codex assisted code development, diagnostic calculations, plotting, language editing, and \LaTeX{} preparation. The retained project-related session metadata identify \texttt{gpt-5.5}, \texttt{gpt-5.6-sol}, \texttt{gpt-5.6-luna}, \texttt{gpt-6-astra}, and \texttt{gpt-6.1-sol}. The October 2026 revision session used the recorded identifier \texttt{gpt-6.1-sol} through the Codex VS Code integration, client version \texttt{0.162.0-alpha.17.2}. These records identify the models and client used in the documented sessions, but do not resolve immutable model snapshots or assign every historical code edit to a model.

Programming assistance covered event parsing, particle and event selections, conditional profiles, moment and uncertainty calculations, tune comparisons, and plotting scripts. Diagnostic work addressed variance conventions, covariance terms in propagated errors, empirical-bootstrap checks, and local and inclusive width normalization. All plotted physics values were calculated from event records or numerical tables using executable scripts.

Validation combined direct data reanalysis with complementary numerical checks. The central analysis processed all 60 million source event records, and a full scan of the six baseline particle samples reproduced the stored charged multiplicities, INEL$>0$-like masks, and summed transverse momenta. Conditional moments and errors were recomputed from integer joint-count tables. Absolute-width and coefficient-of-variation errors were checked with 2,000 empirical-bootstrap replicates in all 148 matched bins. For the inclusive-normalized width, the covariance with the inclusive mean was retained and the influence-function expression was checked by numerical perturbation. Source syntax and compilation checks, small validation-only generator runs, and reproduction of the numerical tables and plots from a fresh S2 extraction provided further checks. S1 and S2 give the detailed validation records and task-to-evidence map.

The validation code was also developed with AI assistance, so these checks establish numerical consistency with retained inputs rather than independent human validation. They do not recover missing historical production settings. Scientific interpretation and responsibility for the manuscript remain with the authors.

\section{Inclusive distributions and motivation for conditioning}
\label{sec:inclusive}

\begin{figure*}[!tbp]\centering
\includegraphics[width=\textwidth]{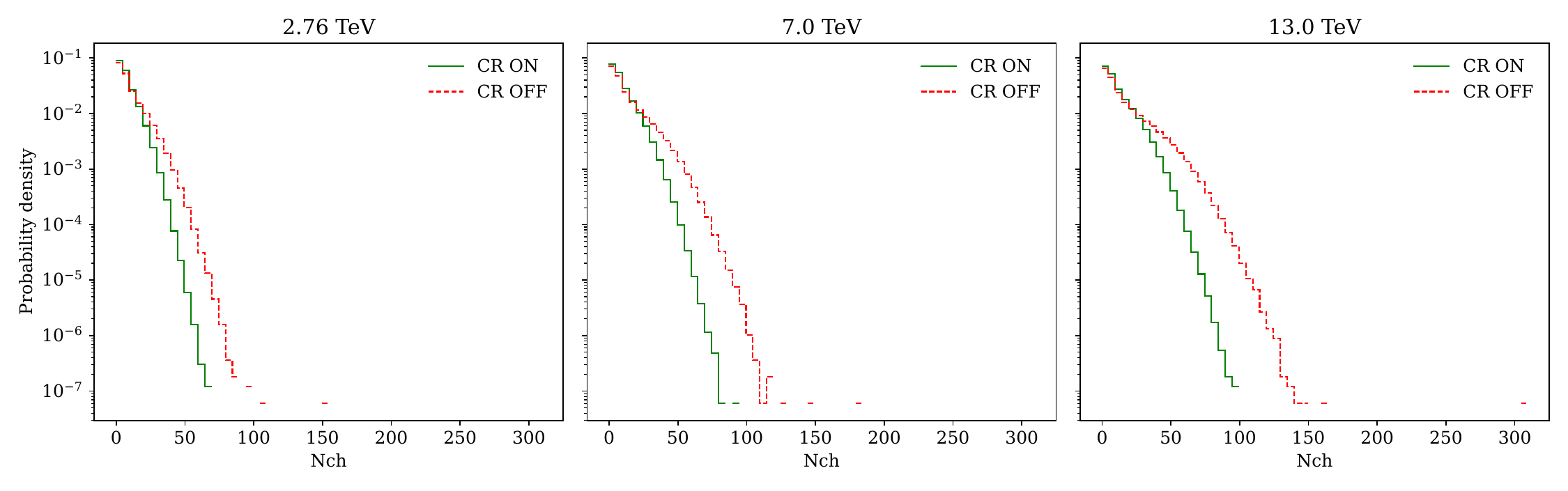}
\caption{\label{fig:inclusive_nch}Inclusive charged-multiplicity distributions for the Monash baseline at $\sqrt{s}=2.76$, 7, and 13~TeV (left to right). Green solid and red dashed step histograms denote CR ON and CR OFF, respectively. Events satisfy the INEL$>0$-like selection without an $n_{\rm MPI}$ cut; charged particles satisfy $|\eta|<0.8$ and $0.15<p_{\rm T}<20$~GeV/$c$. Bin width is 5, and the ordinate is probability per unit multiplicity, normalized to unit integral. Empty bins are omitted on the logarithmic axis; statistical error bars are not displayed.}
\end{figure*}

\begin{figure*}[!tbp]\centering
\includegraphics[width=\textwidth]{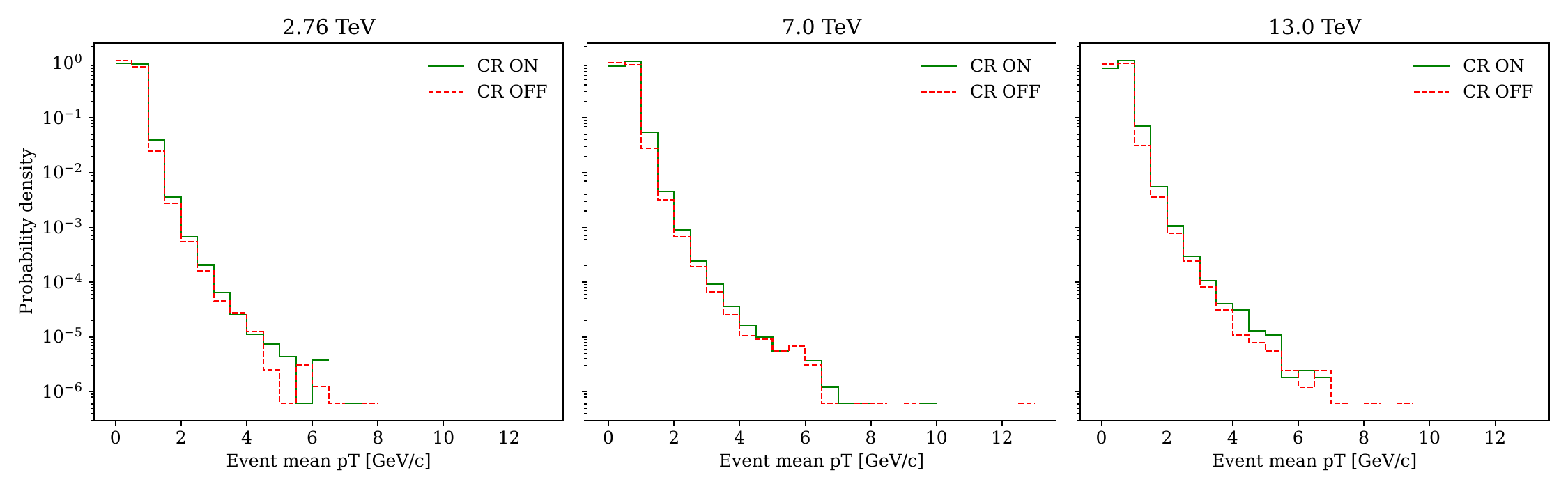}
\caption{\label{fig:inclusive_meanpt}Distribution of event-mean transverse momentum for the same energies, particle acceptance, and CR color convention as Fig.~\ref{fig:inclusive_nch}. The INEL$>0$-like selection is supplemented by $N_{\rm ch}>0$, without an $n_{\rm MPI}$ cut. Each event has equal weight. Bin width is 0.5~GeV/$c$, and each distribution is normalized to unit integral. Empty bins are omitted on the logarithmic axis; statistical error bars are not displayed.}
\end{figure*}

The inclusive multiplicity and event-mean transverse-momentum distributions show the overall CR dependence in the six Monash baseline samples (Figs.~\ref{fig:inclusive_nch} and \ref{fig:inclusive_meanpt}). Events satisfy the INEL$>0$-like selection without an MPI cut. The multiplicity distribution includes $N_{\rm ch}=0$: events can pass the wider event selection while containing no charged particle within the analysis cuts.

For events with $N_{\rm ch}>0$, the event-mean transverse momentum is the arithmetic mean of the accepted particle transverse momenta. Each event receives equal weight. This event-level distribution complements the multiplicity-dependent $\langle p_{\rm T}\rangle(N_{\rm ch})$ profile used in the ALICE comparison. The histograms in each figure have unit integral; their bin widths are specified in the captions. The explicit event-mean definition is given in Supplementary Material S1.

CR lowers the mean charged multiplicity and raises the average event-mean transverse momentum, with larger shifts at higher energy (Table~\ref{tab:inclusive_summary}). It also reduces the high-multiplicity population and increases the fraction of events with large event-mean transverse momentum. The averages give equal weight to events in their respective selected samples.

\begin{table}[!tbp]\centering\footnotesize
\caption{\label{tab:inclusive_summary}Inclusive Monash results for the selections of Figs.~\ref{fig:inclusive_nch} and \ref{fig:inclusive_meanpt}. Each pair is CR ON / OFF. Mean-$p_{\rm T}$ values are in GeV/$c$; tail fractions are percentages.}
{\ifmarkchanges\color{revisionblue}\else\color{black}\fi\begin{tabular}{lrrr}\toprule
$\sqrt{s}$ [TeV]&2.76&7&13\\\midrule
$\langle N_{\rm ch}\rangle$&6.92 / 8.74&8.74 / 11.65&10.25 / 14.14\\
$\langle\langle p_{\rm T}\rangle_{\rm event}\rangle$&0.530 / 0.502&0.558 / 0.515&0.580 / 0.526\\
Multiplicity change [\%]&$-20.8$&$-25.0$&$-27.5$\\
Mean-$p_{\rm T}$ change [\%]&5.7&8.3&10.2\\
$N_{\rm ch}\geq20$ [\%]&--&--&15.8 / 25.3\\
$\langle p_{\rm T}\rangle_{\rm event}\geq1$ [\%]&--&--&3.90 / 1.79\\\bottomrule
\end{tabular}}\end{table}

These inclusive shifts mix contributions from events with different MPI counts. The multiplicity distribution can be written as

\begin{equation}
P(N_{\rm ch})=\sum_m P(N_{\rm ch}\mid n_{\rm MPI}=m)P(n_{\rm MPI}=m),
\end{equation}

Here the probabilities refer to the INEL$>0$-like sample without an $n_{\rm MPI}$ cut, with $N_{\rm ch}=0$ retained. The inclusive distribution combines conditional distributions at different MPI counts, weighted by their event fractions. It therefore does not by itself reveal how CR changes multiplicity dispersion at a fixed MPI count.

To separate these contributions, we compare CR ON and CR OFF at the same integer MPI count. We require $n_{\rm MPI}>2$ and retain common bins with at least 500 events in each sample. Within each matched bin, we measure the mean and the absolute and relative widths.

\section{Conditional response and fluctuation measures}
\label{sec:conditional}

At each selected integer MPI count, $P(N_{\rm ch}\mid n_{\rm MPI}=m)$ describes the within-bin multiplicity distribution. Its moments compare the two CR states without mixing different MPI populations~\cite{Cowan1998,GorensteinGazdzicki2011}.

For the $N_m$ selected events in bin $m$, the conditional mean and absolute standard deviation are

\begin{align}
\mu_m&=\langle N_{\rm ch}\rangle_m=\frac{1}{N_m}\sum_{i:m_i=m}N_{{\rm ch},i},\\
\sigma_m&=\left[\frac{1}{N_m}\sum_{i:m_i=m}(N_{{\rm ch},i}-\mu_m)^2\right]^{1/2}.
\end{align}

Using the divisor $N_m$ gives equal weight to each event and defines the population moments of the selected sample. The standard deviation $\sigma_m$ is the RMS multiplicity residual about $\mu_m$, expressed on the $N_{\rm ch}$ scale.

To interpret an absolute width in relation to its conditional mean, we use the following normalized measures:

\begin{equation}
C_m=\frac{\sigma_m}{\mu_m},\qquad F_m=\frac{\sigma_m^2}{\mu_m}.
\end{equation}

The coefficient of variation $C_m$ expresses the standard deviation as a fraction of the local mean. The scaled variance $F_m$ compares the variance with the mean and equals one for an ideal Poisson distribution; it is commonly used in multiplicity-fluctuation studies~\cite{ALICEMultFluct2021}. The relation $C_m^2=F_m/\mu_m$ allows the two measures to have different CR orderings.

To compare the changes in mean and width, define $R_{\mu,m}=\mu_{m,{\rm ON}}/\mu_{m,{\rm OFF}}$, $R_{\sigma,m}=\sigma_{m,{\rm ON}}/\sigma_{m,{\rm OFF}}$, and $R_{C,m}=C_{m,{\rm ON}}/C_{m,{\rm OFF}}$. Their algebraic relation is

{\ifmarkchanges\color{revisionblue}\else\color{black}\fi\begin{equation}\label{eq:scale_width}
R_{C,m}=\frac{R_{\sigma,m}}{R_{\mu,m}}.
\end{equation}}

When $R_{\mu,m}<R_{\sigma,m}<1$, CR ON has a smaller absolute width but a larger relative width: the mean decreases by a larger fraction than the standard deviation. Equation~\ref{eq:scale_width} expresses this balance between changes in scale and dispersion.

\begin{figure*}[!tbp]\centering\includegraphics[width=\textwidth]{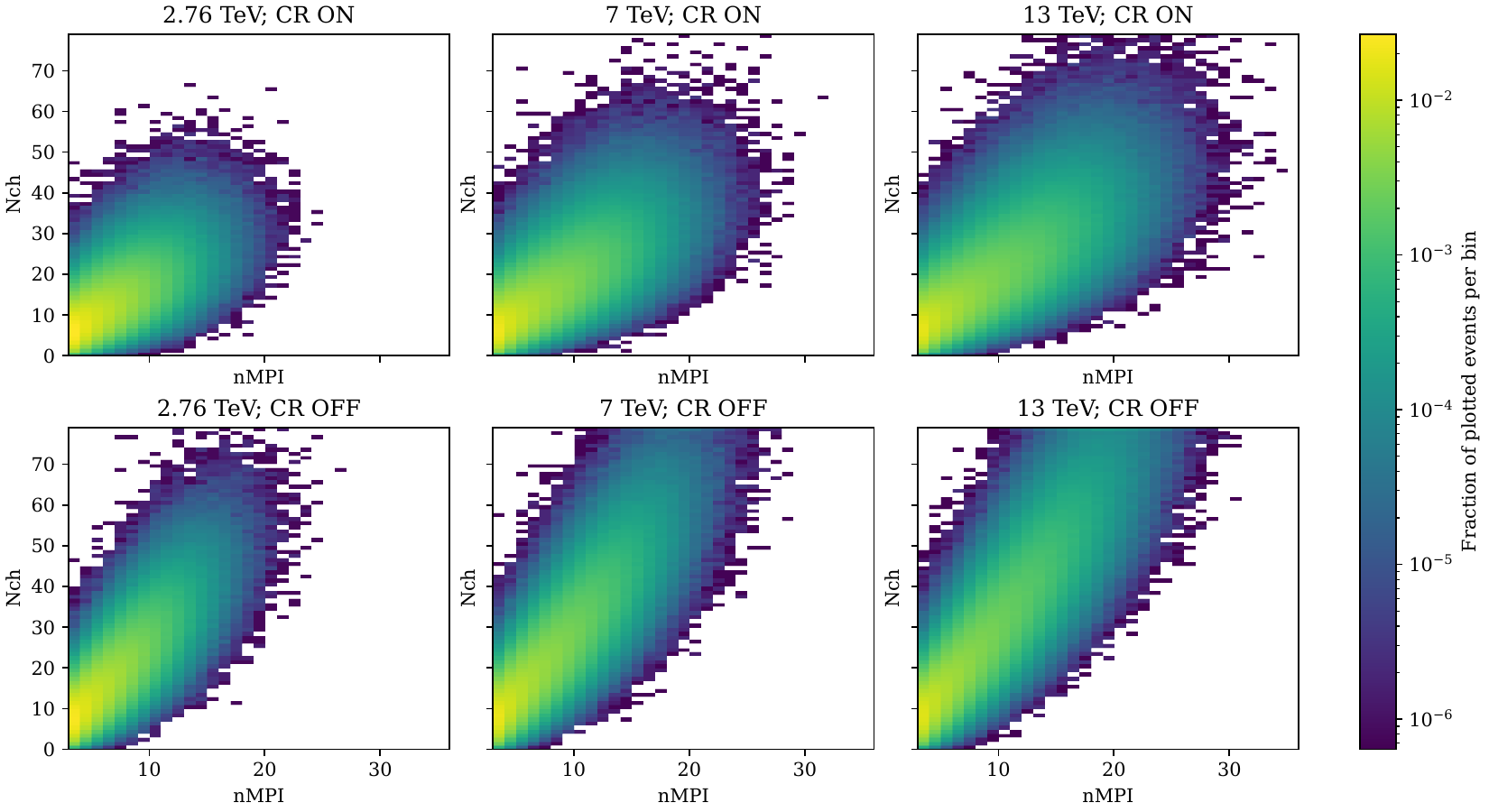}
\caption{\label{fig:density}Normalized event density in the PYTHIA-specific $n_{\rm MPI}$--$N_{\rm ch}$ plane. The color bar gives the fraction of plotted events in each bin, obtained by dividing the bin counts by their sum within each panel. The map characterizes the relation between internal MPI activity and final charged multiplicity in each simulated sample.}
\end{figure*}

\begin{figure*}[!tbp]\centering\includegraphics[width=\textwidth]{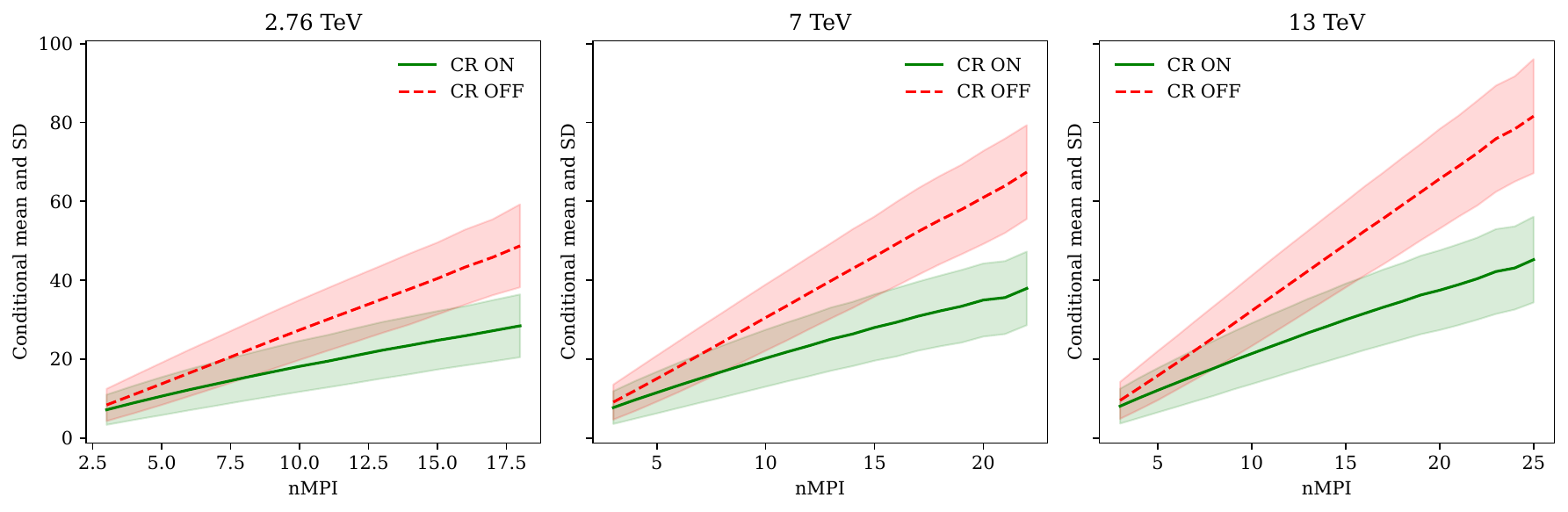}
\caption{\label{fig:profiles}Conditional mean $\mu_m=\langle N_{\rm ch}\rangle_m$ at each retained integer $m=n_{\rm MPI}$. Shaded bands show $\mu_m\pm\sigma_m$, where $\sigma_m$ is the absolute conditional standard deviation. Only bins containing at least 500 events are shown. The absolute width is interpreted together with the normalized measures in Fig.~\ref{fig:normfluc}.}
\end{figure*}

CR OFF shifts the conditional means to larger multiplicities and produces a steeper response. The density maps in Fig.~\ref{fig:density} retain the event spread, while Fig.~\ref{fig:profiles} shows the means with absolute-width bands. Supplementary Material S1 gives the correlation coefficients, fitted slopes, and sample-level width summaries. The main comparison below uses the local widths at matched MPI counts.

\begin{figure*}[!tbp]\centering
\includegraphics[width=0.94\textwidth]{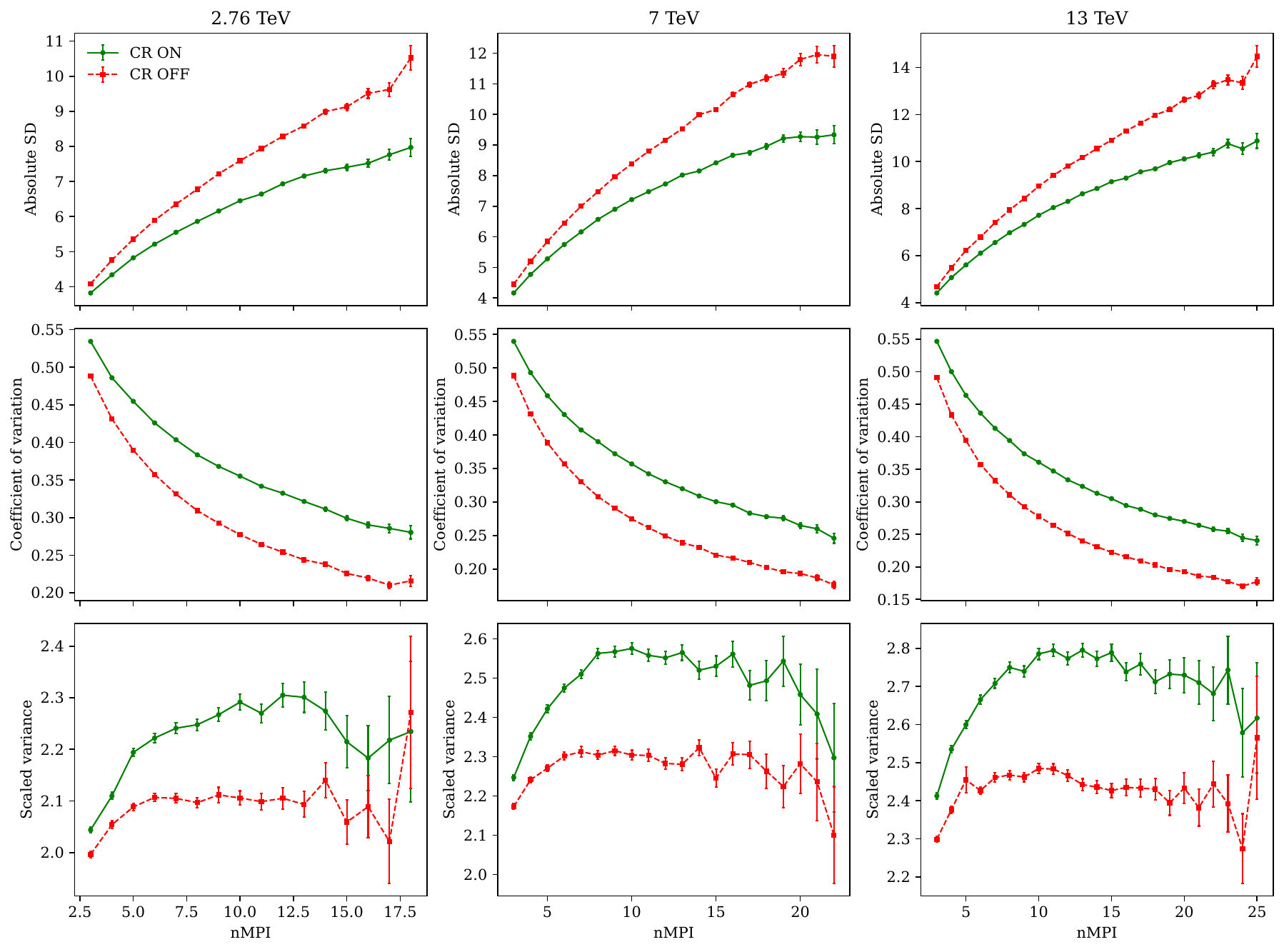}
\caption{\label{fig:normfluc}Absolute width $\sigma_m$ in panels (a)--(c), coefficient of variation $C_m$ in panels (d)--(f), and scaled variance $F_m$ in panels (g)--(i), all versus $n_{\rm MPI}$. Columns correspond to 2.76, 7, and 13~TeV from left to right. Green solid circles denote CR ON; red dashed squares denote CR OFF. Points are binwise estimates and lines only guide the eye. Vertical bars show one-standard-error statistical uncertainties propagated from the second, third, and fourth central moments; most are smaller than the markers. Only bins containing at least 500 events are shown. CR ON has a smaller absolute width but larger coefficient of variation in all 16, 20, and 23 matched bins at 2.76, 7, and 13~TeV, respectively.}
\end{figure*}

The absolute width $\sigma_m$ increases with MPI count but remains smaller with CR ON (Fig.~\ref{fig:normfluc}(a)--(c)). At a common MPI count, the widths increase with collision energy. The CR separation generally grows with activity.

The coefficient of variation $C_m$ decreases with MPI count because the mean grows faster than the absolute width (Fig.~\ref{fig:normfluc}(d)--(f)). CR ON has the larger $C_m$ in every retained matched bin. Its energy dependence is weaker than that of $\sigma_m$.

The scaled variance $F_m$ exceeds the Poisson reference in all displayed baseline bins (Fig.~\ref{fig:normfluc}(g)--(i)). CR ON gives larger values over most of the retained range, although endpoint uncertainties limit individual-bin interpretation. Since $C_m^2=F_m/\mu_m$, a falling coefficient of variation can coexist with variance above the Poisson reference.

Table~\ref{tab:matched_ratios} summarizes the ON/OFF ratios using the smaller ON/OFF event count as the common weight for each matched bin. The averaging procedure is given in Supplementary Material S1.

In all 16, 20 and 23 matched bins at 2.76, 7 and 13~TeV, respectively, $R_{\mu,m}<R_{\sigma,m}<1$ (Table~\ref{tab:matched_ratios}). The fractional decrease in the mean is therefore larger than that in the absolute width, explaining the larger relative width with CR ON.

\begin{table}[!tbp]\centering\small
\caption{\label{tab:matched_ratios}Common-bin-weighted ON/OFF ratios for the baseline, with $w_m=\min(N_{m,{\rm ON}},N_{m,{\rm OFF}})$. Ratios are evaluated bin by bin before averaging; the averages therefore need not obey the binwise identity in Eq.~\ref{eq:scale_width}.}
{\ifmarkchanges\color{revisionblue}\else\color{black}\fi\begin{tabular}{lrrrr}\toprule
$\sqrt{s}$ [TeV]&Bins&$\overline R_\mu$&$\overline R_\sigma$&$\overline R_C$\\\midrule
2.76&16&0.766&0.895&1.174\\
7&20&0.741&0.891&1.210\\
13&23&0.728&0.890&1.232\\\bottomrule
\end{tabular}}\end{table}

\section{Comparison with data}
\label{sec:alice}

\begin{figure}[!t]\centering\includegraphics[width=\columnwidth]{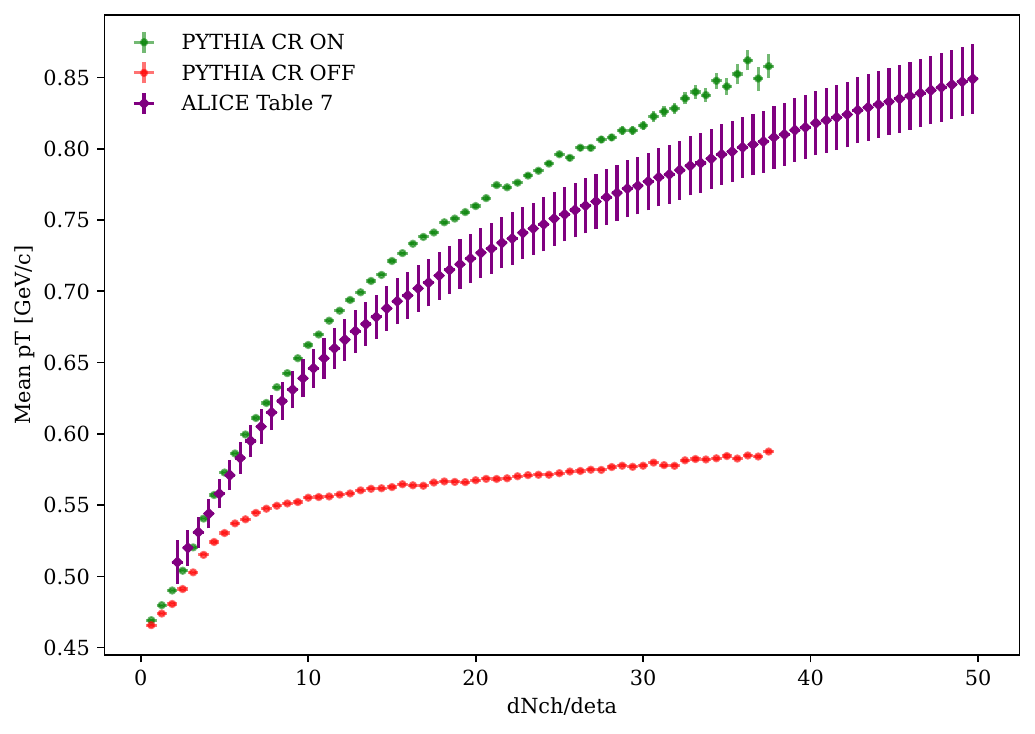}
\caption{\label{fig:alicept}Mean transverse momentum versus charged-particle density. PYTHIA uses the INEL$>0$-like selection without an $n_{\rm MPI}$ cut. Vertical PYTHIA bars are Monte Carlo standard errors of the mean; they do not represent tune, selection, or generator-model uncertainty. ALICE values are from the associated HEPData record~\cite{ALICEPt2019,ALICEHEPDataTable7}.}
\end{figure}

The multiplicity dependence of mean transverse momentum provides an experimental reference for the CR response~\cite{ALICE:2013rdo}. Figure~\ref{fig:alicept} compares $\langle p_{\rm T}\rangle(N_{\rm ch})$ at 13~TeV with the ALICE measurement~\cite{ALICEPt2019}, using numerical values from the associated HEPData table~\cite{ALICEHEPDataTable7}. This comparison does not condition on MPI count.

The CR-ON curve follows the measured trend more closely over much of the displayed range. A quantitative goodness-of-fit comparison would additionally require common particle definitions and the experimental uncertainty correlations.

The stored samples do not reproduce the full ALICE primary-particle definition. The baseline arrays lack the ancestry, status and lifetime information needed to remove secondary particles according to that definition. No correction or systematic uncertainty has been assigned to this difference, and no detector response is applied. The comparison consequently indicates agreement in trend within the stated particle-definition and selection limits.

\section{Final-state event-activity estimators in PYTHIA}
\label{sec:estimators}

Final-state activity classifiers based on multiplicity, spherocity, transverse activity, and flattenicity have been studied extensively~\cite{Ortiz:2015ttf,Martin:2016igp,Prasad:2024gqq,Prasad:2025yfj,Mendez:2025dqz}. Related approaches include MPI-sensitive regression~\cite{Ortiz:2020rwg} and experimental uncorrelated seeds compared with simulated MPI~\cite{ALICE:2013tla}. We examine five simulated estimators to characterize how the conditional response is reflected in different final-state regions.

The five estimators are midrapidity multiplicity, forward multiplicity, transverse-region multiplicity, the transverse-region scalar-$p_{\rm T}$ sum, and forward--backward asymmetry. All use charged particles with $p_{\rm T}>0.15$~GeV/$c$ and no upper momentum cut. The midrapidity and forward acceptances are $|\eta|<0.8$ and $2.5<|\eta|<5.0$, respectively.

The transverse region is defined by $\pi/3<|\Delta\phi|<2\pi/3$ relative to the leading charged particle with $|\eta|<0.8$ and $p_{\rm T}>0.5$~GeV/$c$. Associated particles satisfy $|\eta|<0.8$. Their count and scalar momentum sum define the two transverse observables.

The asymmetry is $A_{\rm FB}=|N_F-N_B|/(N_F+N_B)$, where $N_F$ and $N_B$ are the counts in the positive and negative forward intervals. It is evaluated only for $N_F+N_B>0$.

The baseline estimator analysis requires $n_{\rm MPI}>2$ and positive midrapidity multiplicity without an upper momentum cut. Transverse observables additionally require the leading particle, and asymmetry requires a nonzero denominator. Correlations use the events satisfying the relevant requirements.

\begin{figure*}[p]\centering
\includegraphics[width=\textwidth,height=0.85\textheight,keepaspectratio]{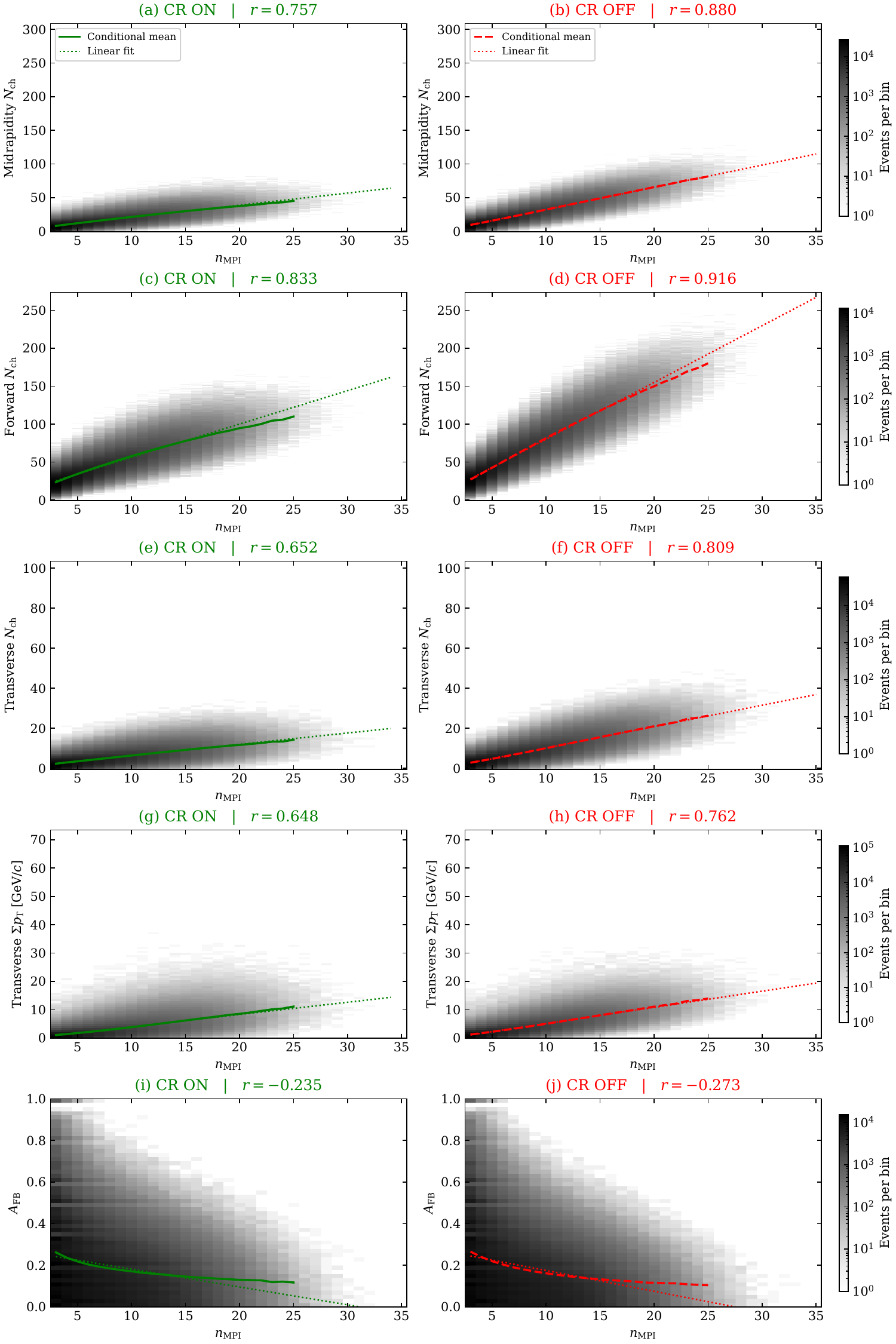}
\caption{\label{fig:proxymaps}Final-state estimators versus $n_{\rm MPI}$ at 13~TeV for CR ON (left) and CR OFF (right). Rows show midrapidity multiplicity, forward multiplicity, transverse multiplicity, transverse scalar-$p_{\rm T}$ sum, and forward--backward asymmetry. Gray shading gives raw events per bin on a logarithmic scale, with common axes and color normalization within each row. These maps show unnormalized counts; the logarithmic scale retains the sparsely populated tails. Green solid and red dashed curves show the CR-ON and CR-OFF conditional means for activity bins containing at least 500 events; dotted lines show event-level least-squares fits. The full selected event ranges are retained. Each panel reports Pearson's $r$ evaluated from the individual events, including activity bins below the profile-display threshold.}
\end{figure*}

Fig.~\ref{fig:proxymaps} shows the event distributions, conditional means, and linear fits. Means are displayed for bins with at least 500 events; Pearson coefficients and event-level fits use all selected events, including bins below that threshold. Table~\ref{tab:allproxies} lists the coefficients.

\begin{table}[!tbp]\centering\small
\caption{\label{tab:allproxies}Pearson association with PYTHIA $n_{\rm MPI}$ at 13~TeV for all five estimators.}
\begin{tabular}{lrr}\toprule Estimator&CR ON&CR OFF\\\midrule
Midrapidity $N_{\rm ch}$&0.757&0.880\\
Forward $N_{\rm ch}$&0.833&0.916\\
Transverse $N_{\rm ch}$&0.652&0.809\\
Transverse $\Sigma p_{\rm T}$&0.648&0.762\\
$A_{\rm FB}$&$-0.235$&$-0.273$\\\bottomrule
\end{tabular}\end{table}

Forward multiplicity has the strongest positive linear association among these estimators under their respective selections, followed by midrapidity multiplicity and the transverse observables (Table~\ref{tab:allproxies}). All four have rising conditional means and larger correlations with CR OFF.

Each coefficient describes the estimator under its own selection. Radiation, beam remnants, hadronization, decays, hard scattering, and acceptance also shape its response, so correlations alone give an incomplete measure of MPI resolution and experimental selection bias.

The asymmetry behaves differently: its conditional mean decreases with MPI count, with modest negative correlations in both CR states. Its curved mean response and broad distributions illustrate the limits of a single linear coefficient for describing the relation to MPI activity.

Low-, mid- and high-activity classes are defined by the one-third and two-thirds quantiles of forward multiplicity, transverse multiplicity, and transverse scalar-$p_{\rm T}$ sum. Quantiles are calculated separately for each CR state, so corresponding labels need not have identical boundaries. Discrete ties can give unequal populations; boundaries and counts are supplied in Supplementary Material S1.

Figure~\ref{fig:proxyclasses} shows the mean MPI count in each activity class. Supplementary Material S1 lists the class boundaries, populations, means, and standard errors. The errors are smaller than the plotting symbols and quantify the precision of the means, rather than the event spread or model uncertainty.

For all three estimators, mean MPI activity rises from the low- to the high-activity class in both CR states. This shows that the selected final-state observables separate populations with different average generator activity, complementing the event-level correlations.

\begin{figure*}[!tbp]
\centering
\includegraphics[width=\textwidth]{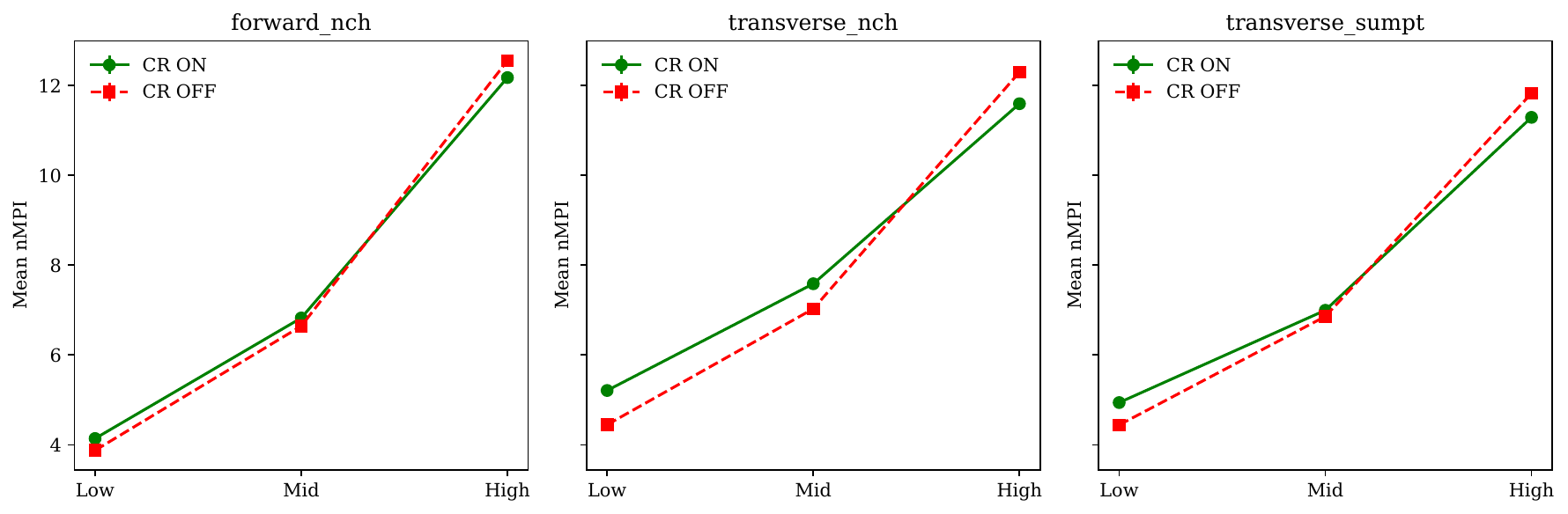}
\captionof{figure}{\label{fig:proxyclasses}Mean PYTHIA $n_{\rm MPI}$ in low-, mid-, and high-activity classes at 13~TeV for three activity estimators: forward multiplicity, transverse-region multiplicity, and transverse-region scalar-$p_{\rm T}$ sum. Supplementary Material S1 lists the plotted class means and their statistical uncertainties. Correlations for all five estimators, including midrapidity multiplicity and $A_{\rm FB}$, are given separately in Table~\ref{tab:allproxies}. Standard errors of the mean are included, but are smaller than the plotting symbols and therefore not visible at this scale.}

\end{figure*}

\section{Tune and analysis-choice sensitivity}
\label{sec:tunes}

The independent Monash, A14, and CP5 samples test the configuration dependence of the conditional result. Within each tune, disabling CR leaves the remaining physics settings unchanged (Sec.~\ref{sec:simulation}).

The smaller absolute width and larger coefficient of variation with CR ON persist in all common retained bins: 23 for Monash, 24 for A14, and 42 for CP5. Table~\ref{tab:tune_fluctuations} gives the event-weighted $C_m$ and $F_m$ summaries. The size of the effect varies among tunes.

The scaled variance has a different tune dependence. Its event-weighted value is larger with CR ON for Monash and A14, whereas CP5 gives nearly equal values, with CR ON slightly smaller. The common ordering of absolute width and coefficient of variation thus does not imply a common ordering of scaled variance.

The tune summaries use each sample's event counts as bin weights and therefore differ from the common-bin ratios in Sec.~\ref{sec:conditional}. Both baseline and tune widths use the population divisor $N_m$, with the weighting procedure specified in S1.

The tune scan also compares forward multiplicity, transverse multiplicity, and the transverse scalar-$p_{\rm T}$ sum. It requires $n_{\rm MPI}>2$ and positive central multiplicity within $0.15<p_{\rm T}<20$~GeV/$c$. The estimators themselves have no upper momentum cut, and the transverse observables also require a leading particle. The additional central momentum bound and the independent samples distinguish this comparison from the baseline estimator analysis in Sec.~\ref{sec:estimators}.

\begin{figure*}[!tbp]\centering\includegraphics[width=\textwidth]{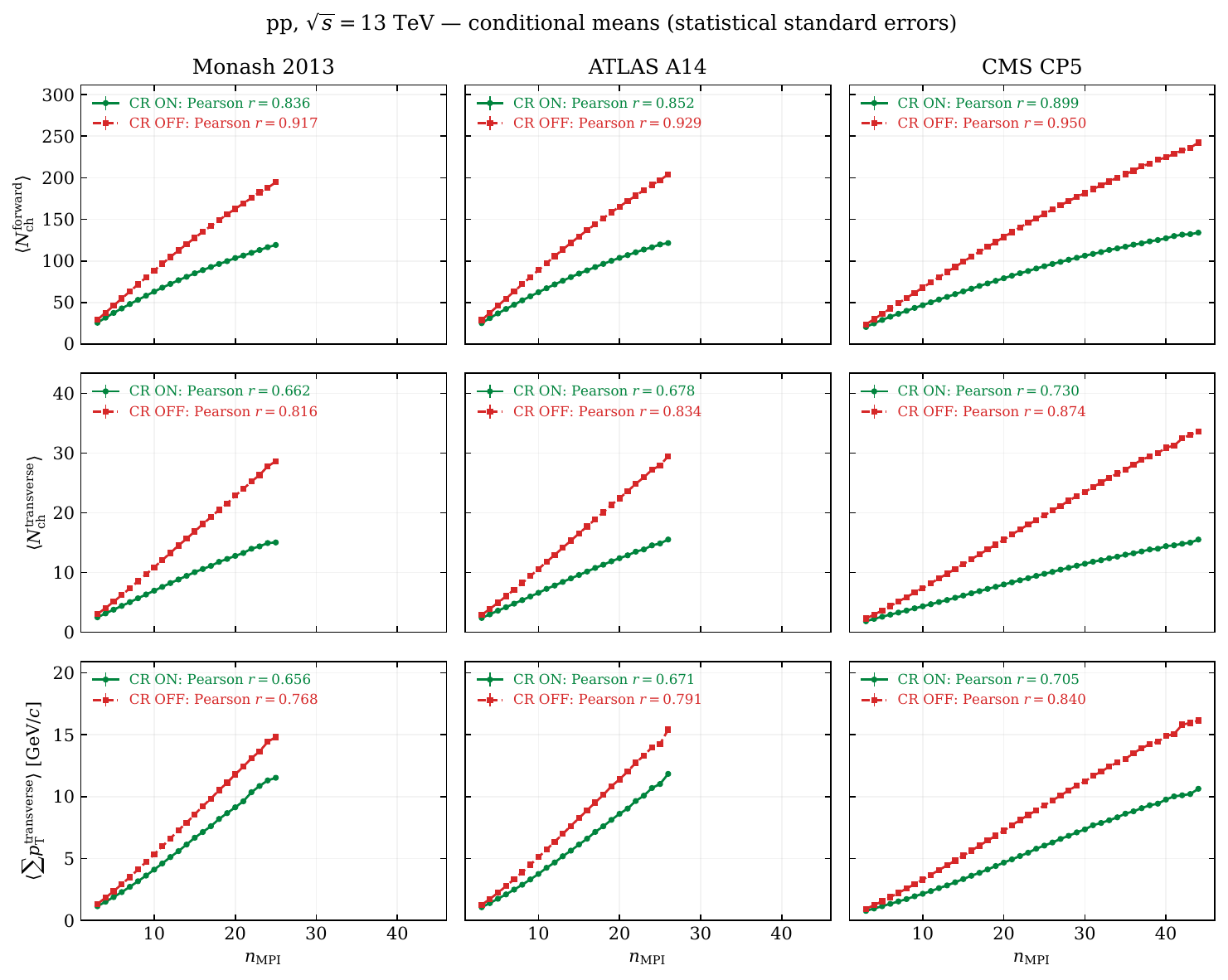}
\caption{\label{fig:tuneproxy}Conditional mean estimators versus $n_{\rm MPI}$ at 13~TeV. Rows show forward multiplicity, transverse multiplicity, and transverse scalar-$p_{\rm T}$ sum; columns show Monash 2013, ATLAS A14, and CMS CP5. Green solid curves with circles denote CR ON; red dashed curves with squares denote CR OFF. Error bars give statistical standard errors of the means, often smaller than the markers. Only integer $n_{\rm MPI}$ bins containing at least 500 selected events are shown; lines guide the eye. Each tune and CR state uses five million generated events with the estimator selections described in the text. Axes are common within each row. These uncertainties describe the means, not the event-by-event spread. Each panel lists the event-level Pearson coefficients for CR ON and CR OFF, calculated from all selected events, including bins below the profile-display threshold.}
\end{figure*}

The three estimators retain the same ranking for all tunes and have larger correlations with CR OFF (Table~\ref{tab:tuneproxy}). Fig.~\ref{fig:tuneproxy} shows their mean responses, with statistical errors of the means for bins containing at least 500 events. Correlations use all selected events.

\begin{table*}[!tbp]\centering
\caption{\label{tab:tuneproxy}Equal-statistics 5M-event tune comparison of estimator--$n_{\rm MPI}$ Pearson coefficients.}\begin{tabular}{llcc}
\toprule
Proxy & Tune & CR ON & CR OFF \\
\midrule
Forward multiplicity & Monash & 0.836 & 0.917 \\
Forward multiplicity & ATLAS A14 & 0.852 & 0.929 \\
Forward multiplicity & CMS CP5 & 0.899 & 0.950 \\
\addlinespace
Transverse multiplicity & Monash & 0.662 & 0.816 \\
Transverse multiplicity & ATLAS A14 & 0.678 & 0.834 \\
Transverse multiplicity & CMS CP5 & 0.730 & 0.874 \\
\addlinespace
Transverse scalar-$p_{\rm T}$ sum & Monash & 0.656 & 0.768 \\
Transverse scalar-$p_{\rm T}$ sum & ATLAS A14 & 0.671 & 0.791 \\
Transverse scalar-$p_{\rm T}$ sum & CMS CP5 & 0.705 & 0.840 \\
\bottomrule
\end{tabular}

\end{table*}

\begin{table}[!tbp]\centering\small
\caption{\label{tab:tune_fluctuations}13~TeV conditional-fluctuation summaries for $n_{\rm MPI}>2$, retaining bins with at least 500 events. Each numerical pair is CR ON / OFF. The last column lists bins with $C_{m,{\rm ON}}>C_{m,{\rm OFF}}$ / all matched bins.}
{\ifmarkchanges\color{revisionblue}\else\color{black}\fi\begin{tabular}{lrrr}\toprule
Configuration&Weighted $C_m$&Weighted $F_m$&Bins\\\midrule
Monash&0.420 / 0.350&2.808 / 2.575&23 / 23\\
A14&0.414 / 0.339&2.675 / 2.416&24 / 24\\
CMS CP5&0.436 / 0.358&2.595 / 2.597&42 / 42\\\bottomrule
\end{tabular}}\end{table}

The MPI selection and fitted range change the Monash slopes by several percent, while an additional positive-central-multiplicity requirement changes them by less than 0.2\%. Supplementary Material S1 gives the numerical comparisons and distinguishes event-level fits from conditional-profile fits. These checks describe analysis sensitivity separately from statistical errors and do not constitute a complete model-uncertainty estimate.

\section{Discussion and limitations}
\label{sec:discussion}

\subsection{Effect of normalization on conditional widths}
\label{sec:weber_comparison}
The relative size of conditional fluctuations depends on the reference scale. Alongside normalization by the local conditional mean, one can express the width relative to the inclusive multiplicity of each CR state. Let $G_s=\langle N_{\rm ch}\rangle_s$ denote that inclusive mean for the INEL$>0$-like sample, without an MPI cut and with $N_{\rm ch}=0$ retained. Define
\begin{equation}
W_{m,s}=\frac{\sigma_{m,s}}{G_s},\qquad
R_G=\frac{G_{\rm ON}}{G_{\rm OFF}}.
\label{eq:weber_global_width}
\end{equation}
The CR ON/OFF ratios then satisfy
\begin{equation}
R_{W,m}=\frac{R_{\sigma,m}}{R_G},\qquad
R_{C,m}=\frac{R_{\sigma,m}}{R_{\mu,m}}.
\label{eq:weber_normalization_ratios}
\end{equation}
Separate inclusive normalization gives $R_{W,m}>1$ when $R_G<R_{\sigma,m}<1$, whereas normalization by the conditional mean gives $R_{C,m}>1$ when $R_{\mu,m}<R_{\sigma,m}<1$. Thus the two relative widths probe different changes in scale. All comparisons here match integer $m$; matching the coordinate $m/\langle n_{\rm MPI}\rangle_s$ would generally select different MPI counts in the two CR states. A shared fixed normalization would preserve the absolute-width ordering.

Table~\ref{tab:weber_normalizations} summarizes the three width definitions across all 148 matched bins. In the 13~TeV baseline, absolute narrowing and inclusive-normalized broadening coexist in all 23 bins. For example, at $m=10$, $R_{\mu,m}=0.663334$, $R_{\sigma,m}=0.862401$, and $R_G=0.724866$, giving $R_{C,m}=1.300100$ and $R_{W,m}=1.189738\pm0.004710$, where the uncertainty is one statistical standard error. The same central absolute-versus-inclusive-normalized ordering holds in every retained bin of the independent 13~TeV tune comparisons. At 2.76~TeV, $R_{W,m}>1$ holds in 14 of 16 bins, whereas $R_{C,m}>1$ holds in all 16; inclusive and local normalization are therefore not interchangeable.

\begin{table*}[!tbp]\centering\small
\caption{\label{tab:weber_normalizations}Absolute and relative conditional-width orderings at matched integer MPI counts. $B$ is the number of retained matched bins; the three count columns give bins with absolute narrowing, local-relative broadening, and inclusive-relative broadening. Parentheses in the final column give bins whose marginal normal-approximation 95\% lower bound on $R_{W,m}$ exceeds unity. The intervals are marginal rather than simultaneous.}
\begin{tabular}{lrrrrrr}\toprule
Sample & $m$ range & $R_G$ & $B$ & $R_{\sigma,m}<1$ & $R_{C,m}>1$ & $R_{W,m}>1$\\\midrule
Baseline, 2.76 TeV & 3--18 & 0.791969 & 16 & 16 & 16 & 14 (12)\\
Baseline, 7 TeV & 3--22 & 0.749935 & 20 & 20 & 20 & 20 (18)\\
Baseline, 13 TeV & 3--25 & 0.724866 & 23 & 23 & 23 & 23 (22)\\
Independent Monash, 13 TeV & 3--25 & 0.728878 & 23 & 23 & 23 & 23 (22)\\
Independent A14, 13 TeV & 3--26 & 0.707613 & 24 & 24 & 24 & 24 (23)\\
Independent CP5, 13 TeV & 3--44 & 0.615997 & 42 & 42 & 42 & 42 (40)\\\bottomrule
\end{tabular}
\end{table*}

Statistical errors for $W_{m,s}$ use empirical influence-function propagation, including covariance between the conditional standard deviation and the estimated inclusive mean. The $C_{m,s}$ errors include the conditional mean--variance covariance. Ratios treat the CR samples as independent, subject to the historical seed-record limitation. The variance divisor is the bin event count, as in Sec.~\ref{sec:conditional}. S1 gives the propagation expressions and every-bin results; S2 supplies the corresponding numerical tables and code.

This scale dependence helps interpret earlier studies of conditional fluctuations. Weber et al.~\cite{Weber:2018ddv} discuss conditional broadening alongside self-normalized MPI--multiplicity relations. The present results show how such broadening can coexist with a smaller absolute width: the inclusive mean or the local conditional mean decreases more strongly than the conditional standard deviation. Their discussion does not specify numerical conditional standard deviations or their normalization, so the width descriptions cannot be equated unambiguously. Moreover, their PYTHIA~8.230 non-diffractive samples use ALICE-defined primary charged particles in $|\eta|<1$, while our samples use the definitions in Sec.~\ref{sec:selections}. The normalization comparison therefore explains a compatible pattern on the same events, without establishing direct numerical agreement between the two simulations.

Across the configurations studied here, the measured ordering $R_{\mu,m}<R_{\sigma,m}<1$ separates mean suppression from absolute-width suppression. It quantifies why local relative fluctuations increase despite absolute narrowing, with matched-bin uncertainties and checks across energies and tunes. The persistence of this pattern characterizes the tested configurations; its generality requires further model comparisons.

The conditional correlations and the ALICE mean-$p_{\rm T}$ comparison provide complementary views of the final state. CR OFF produces steeper responses and stronger associations with internal MPI activity, whereas CR ON follows the displayed ALICE mean-$p_{\rm T}$ trend more closely within the limitations of Sec.~\ref{sec:alice}. Correlation strength alone is therefore insufficient to judge experimental agreement. The estimator results complement established classifier studies~\cite{Prasad:2024gqq,Prasad:2025yfj}.

The event and particle definitions, MPI selection, bin retention, and fit range set the scope of these results. A closer experimental comparison requires the full primary-particle information and detector response.

The mapping from MPI activity to final particles depends on shower and hadronization models. PYTHIA uses string fragmentation, whereas Herwig uses a cluster model; MPI and CR are not exclusive to PYTHIA. Cross-generator comparisons~\cite{Ortiz:2020rwg} and separately optimized CR-OFF configurations are needed to assess the generality of the observed ordering.

\section{Conclusions}
\label{sec:conclusions}

For the tested PYTHIA configurations, CR lowers both the conditional mean multiplicity and its absolute width at fixed $n_{\rm MPI}$. The larger fractional decrease in the mean produces a higher coefficient of variation with CR ON in every retained matched bin at 2.76, 7, and 13~TeV.

The same absolute-versus-relative width ordering is found in independent Monash, A14, and CMS CP5 samples. Its magnitude depends on the tune and analysis choices, while the scaled variance shows a different configuration dependence.

Comparisons with other generators, separately optimized CR-OFF configurations, and samples with complete primary-particle information would establish how this conditional pattern extends beyond the present setup.

\section*{Declaration of generative AI and AI-assisted technologies in the manuscript preparation process}
OpenAI Codex assisted language editing, \LaTeX{} preparation, analysis-code development, diagnostic calculations, and data-derived plotting. Section~\ref{sec:ai_methods} identifies the recorded models and client version, the assisted tasks, the executable validation performed, and the limits of the historical records. The authors take responsibility for the scientific content, interpretation, and final submitted version. No AI system is listed as an author.

\section*{Data and code availability}
Supplementary Material S1 gives the analysis definitions, numerical results, matched-bin uncertainties, and production-provenance limitations. Supplementary Material S2 contains the analysis and plotting code, integer conditional joint-count inputs, numerical tables, available generation configurations and production records, and the AI-assistance evidence map. Running \texttt{bash RUN\_ANALYSIS.sh} reproduces the central conditional results, plots, and normalization comparison from the supplied inputs. The large retained event and particle files are available from the corresponding author upon reasonable request. Missing historical seed and effective lifetime/decay records limit reproducibility to reanalysis of the retained samples rather than exact historical generator replay.

\section*{Declaration of competing interest}
The authors declare that they have no known competing financial interests or personal relationships that could have appeared to influence the work reported in this paper.

\section*{Funding}
This research did not receive any specific grant from funding agencies in the public, commercial, or not-for-profit sectors.

\begingroup
\let\originalbibitem\bibitem
\renewcommand{\bibitem}[1]{\color{black}\originalbibitem{#1}\csname revisionbibcolor@#1\endcsname}
\expandafter\def\csname revisionbibcolor@SjostrandVanZijl1987\endcsname{\ifmarkchanges\color{revisionblue}\else\color{black}\fi}
\expandafter\def\csname revisionbibcolor@Pythia8\endcsname{\ifmarkchanges\color{revisionblue}\else\color{black}\fi}
\expandafter\def\csname revisionbibcolor@Pythia82\endcsname{\ifmarkchanges\color{revisionblue}\else\color{black}\fi}
\expandafter\def\csname revisionbibcolor@Pythia64\endcsname{\ifmarkchanges\color{revisionblue}\else\color{black}\fi}
\expandafter\def\csname revisionbibcolor@ALICEMult8TeV\endcsname{\ifmarkchanges\color{revisionblue}\else\color{black}\fi}
\expandafter\def\csname revisionbibcolor@ALICEMult13TeV\endcsname{\ifmarkchanges\color{revisionblue}\else\color{black}\fi}
\expandafter\def\csname revisionbibcolor@OrtizVelasquez2013\endcsname{\ifmarkchanges\color{revisionblue}\else\color{black}\fi}
\expandafter\def\csname revisionbibcolor@Cuautle2019FBCorr\endcsname{\ifmarkchanges\color{revisionblue}\else\color{black}\fi}
\expandafter\def\csname revisionbibcolor@Pearson1895\endcsname{\ifmarkchanges\color{revisionblue}\else\color{black}\fi}
\expandafter\def\csname revisionbibcolor@Spearman1904\endcsname{\ifmarkchanges\color{revisionblue}\else\color{black}\fi}
\expandafter\def\csname revisionbibcolor@GorensteinGazdzicki2011\endcsname{\ifmarkchanges\color{revisionblue}\else\color{black}\fi}
\expandafter\def\csname revisionbibcolor@Weber:2018ddv\endcsname{\ifmarkchanges\color{revisionblue}\else\color{black}\fi}
\expandafter\def\csname revisionbibcolor@Ortiz:2015ttf\endcsname{\ifmarkchanges\color{revisionblue}\else\color{black}\fi}
\expandafter\def\csname revisionbibcolor@ALICE:2013tla\endcsname{\ifmarkchanges\color{revisionblue}\else\color{black}\fi}
\expandafter\def\csname revisionbibcolor@ALICE:2013rdo\endcsname{\ifmarkchanges\color{revisionblue}\else\color{black}\fi}
\expandafter\def\csname revisionbibcolor@Ortiz:2020rwg\endcsname{\ifmarkchanges\color{revisionblue}\else\color{black}\fi}
\expandafter\def\csname revisionbibcolor@Martin:2016igp\endcsname{\ifmarkchanges\color{revisionblue}\else\color{black}\fi}
\expandafter\def\csname revisionbibcolor@Mendez:2025dqz\endcsname{\ifmarkchanges\color{revisionblue}\else\color{black}\fi}
\expandafter\def\csname revisionbibcolor@Prasad:2025yfj\endcsname{\ifmarkchanges\color{revisionblue}\else\color{black}\fi}
\expandafter\def\csname revisionbibcolor@Prasad:2024gqq\endcsname{\ifmarkchanges\color{revisionblue}\else\color{black}\fi}
\expandafter\def\csname revisionbibcolor@Bierlich:2022pfr\endcsname{\ifmarkchanges\color{revisionblue}\else\color{black}\fi}
\expandafter\def\csname revisionbibcolor@LHAPDF6\endcsname{\ifmarkchanges\color{revisionblue}\else\color{black}\fi}
\expandafter\def\csname revisionbibcolor@CMSCP5SettingsCard\endcsname{\ifmarkchanges\color{revisionblue}\else\color{black}\fi}
\bibliographystyle{elsarticle-num}\bibliography{references}\endgroup
\end{document}


\maketitle
{\color{black}\textbf{Resubmission note (21 October 2026).} The approximate third configuration has been replaced by fresh CMS CP5 samples using the full-precision CMS tune card and the external NNPDF3.1 NNLO central PDF. All affected fluctuation profiles, matched-bin checks and estimator coefficients have been recalculated from five million events per CR state. Production-time settings, particle data and seeds are retained in \texttt{cp5\_exact/}. The historical production-time seed and effective lifetime/decay records were not retained; exact historical replay is unavailable. The present resubmission draft adds checked integer joint-count inputs, a common conditional-width convention and complete matched-bin statistical uncertainties. The production-provenance limits are stated explicitly below.}

\begin{abstract}
This standalone supplement records the analysis definitions,
calculation sequence, statistical estimators, uncertainty propagation,
selection checks, tune comparisons, and numerical tables underlying manuscript
NPB-D-26-01492R2. It is intended to make explicit details that cannot be retained
in the journal article because of space and presentation constraints. It does
not introduce a separate physics claim; the interpretation and limitations are
the same as in the revised manuscript.
\end{abstract}

\tableofcontents

\section{Purpose and scope}
The analysis asks how color reconnection (CR) changes the conditional relation
between the internal PYTHIA interaction count $\nmpi$ and the final-state
charged-particle multiplicity $\nch$. The two configurations are denoted CR ON
and CR OFF. The intended CR variation disables \texttt{ColourReconnection:reconnect} while retaining the other tune parameters. Complete effective-setting equality is verified directly for corrected CP5. Historical baseline settings are supported by candidate source records rather than certified per-run dumps. No CR-OFF retuning is performed. Consequently, this is a
controlled generator variation, not a comparison of two independently tuned
descriptions of data.

The value $\nmpi$ is returned by \texttt{pythia.info.nMPI()}. It depends on the
PYTHIA version, tune, multiparton-interaction model, regularization, matter
profile, process selection, and internal counting convention. It is not an
experimental observable and is not interpreted as a universal number of
physical scatterings.

\section{Generated samples and analysis selections}
The historical baseline consists of six retained five-million-event samples, described in the manuscript records as PYTHIA 8.316 \texttt{SoftQCD:all} with Monash 2013 at $\sqrt{s}=2.76$, 7 and 13 TeV in each CR state. Complete per-run version/build and effective-setting records are not certified; the retained source family and the provenance limits are documented below.
Equal-statistics samples at 13 TeV are also analysed for Monash 2013, ATLAS A14,
and CMS CP5.

The analysis multiplicity is
\begin{equation}
\nch=\#\{\text{charged final-state rows}:|\eta|<0.8,\quad
0.15<\pt<20~\text{GeV}/c\}.
\end{equation}
An INEL$>0$-like event contains at least one charged final-state particle in
$|\eta|<1$. The generator-level conditional study additionally requires
$\nmpi>2$. A conditional bin is displayed only if it contains at least 500
selected events.

The stored arrays contain final-state identity, charge, and kinematics but not
the complete ancestry, status, and lifetime provenance required to reproduce the exact
ALICE primary-charged-particle definition. No detector simulation, tracking
efficiency, momentum resolution, or smearing is applied. The ALICE-facing
$\langle\pt\rangle(\nch)$ curve uses the INEL$>0$-like and $\nch>0$
requirements without any $\nmpi$ selection.

The tune-generation code selects charged final-state particles with \texttt{isFinal()} and \texttt{isCharged()}. The accepted multiplicity is counted after applying the analysis cuts; the implementation details are retained in S2.

\section{Event-level quantities and bookkeeping}
For each event $i$:
\begin{itemize}
\item $m_i$ is its integer PYTHIA $\nmpi$ value;
\item $N_{{\rm ch},i}$ is its accepted charged-particle count;
\item $N_m$ is the number of selected events satisfying $m_i=m$;
\item $N=\sum_m N_m$ is the total number of selected events in one energy and
CR sample.
\end{itemize}
The conditional distribution
$P(\nch\mid\nmpi=m)$ contains only events having the same $m$. Conditioning
removes changes in the sampled $\nmpi$ distribution from the within-bin
multiplicity dispersion.

\section{Conditional moments}
For a fixed integer $m$, the conditional mean is
\begin{equation}
\mu_m=\langle\nch\rangle_m=\frac{1}{N_m}
\sum_{i:m_i=m}N_{{\rm ch},i}.
\end{equation}
The residual for event $i$ is
$d_i=N_{{\rm ch},i}-\mu_m$. The $k$th population central moment is
\begin{equation}
\mu_{k,m}=\frac{1}{N_m}\sum_{i:m_i=m}d_i^k.
\end{equation}
The conditional variance and absolute conditional standard deviation are
\begin{equation}
V_m=\mu_{2,m},\qquad \sigma_m=\sqrt{V_m}.
\end{equation}
The corrected conditional-width tables consistently use divisor $N_m$ in the baseline and independent tune samples. The earlier tune tables used $N_m-1$; the corrected tune widths equal those earlier widths multiplied by $\sqrt{(N_m-1)/N_m}$. Original values were reproduced before this documented correction. The calculation characterizes
the generated Monte Carlo population. The absolute width $\sigma_m$ is measured
on the same multiplicity scale as $\nch$.

\section{Global ridge width}
The global ridge width is the RMS of all event residuals after subtracting the
conditional mean appropriate to each event:
\begin{align}
\sigma_{\rm ridge}
&=\left[\frac{1}{N}\sum_i
(N_{{\rm ch},i}-\mu_{m_i})^2\right]^{1/2},\\
\sigma_{\rm ridge}^2
&=\sum_m\frac{N_m}{N}\sigma_m^2.
\end{align}
Thus, it is an event-weighted combination of conditional variances, not the
unconditional standard deviation of the full $\nch$ distribution. ``Ridge''
refers to the populated band in the $\nmpi$--$\nch$ response map and not to a
two-particle angular-correlation ridge.

\section{Normalized fluctuation measures}
Since a width commonly increases with its mean, $\sigma_m$ alone cannot show
whether dispersion is larger relative to the local multiplicity scale. Two
normalizations are used:
\begin{align}
C_m&=\frac{\sigma_m}{\mu_m},&
F_m&=\frac{\sigma_m^2}{\mu_m}.
\end{align}
$C_m$ is the conditional coefficient of variation. It is dimensionless and
reports the standard deviation as a fraction of the conditional mean. For
example, $C_m=0.40$ means $\sigma_m=0.40\mu_m$. $F_m$ is the conditional scaled
variance. For an ideal Poisson multiplicity distribution, variance equals mean
and $F_m=1$. A departure from one quantifies non-Poisson dispersion but does not
identify its physical origin.

For a compact sample-level summary, the manuscript also gives
\begin{equation}
C_{\rm glob}=\frac{\sigma_{\rm ridge}}{\langle\nch\rangle},\qquad
F_{\rm glob}=\frac{\sigma_{\rm ridge}^2}{\langle\nch\rangle},
\end{equation}
where $\langle\nch\rangle=N^{-1}\sum_iN_{{\rm ch},i}$. These are not averages
of $C_m$ or $F_m$ and must not be confused with the matched-bin ratios below.

\section{Matched-bin CR comparison}
A matched bin is the same integer $m=\nmpi$ in the CR-ON and CR-OFF samples,
with at least 500 events in each. For every matched bin,
\begin{align}
R_{\mu,m}&=\frac{\mu_{m,{\rm ON}}}{\mu_{m,{\rm OFF}}},&
R_{\sigma,m}&=\frac{\sigma_{m,{\rm ON}}}{\sigma_{m,{\rm OFF}}},&
R_{C,m}&=\frac{C_{m,{\rm ON}}}{C_{m,{\rm OFF}}},&
R_{F,m}&=\frac{F_{m,{\rm ON}}}{F_{m,{\rm OFF}}}.
\end{align}
The common-bin weight is the smaller of the two populations,
\begin{equation}
w_m=\min(N_{m,{\rm ON}},N_{m,{\rm OFF}}),
\end{equation}
so the weight is not determined by events available in only one configuration.
For $X\in\{\mu,\sigma,C,F\}$, the reported energy-level ratio is
\begin{equation}
\overline R_X=\frac{\sum_mw_mR_{X,m}}{\sum_mw_m}.
\end{equation}
$\overline R_\sigma<1$ means a smaller absolute width with CR ON;
$\overline R_C>1$ means a larger width relative to the CR-ON conditional mean.
It is the coexistence of these two inequalities that produces the central
scale--relative-fluctuation separation.

\section{Statistical uncertainties}
For a bin containing $n=N_m$ events, the large-sample moment estimates used by
the analysis are
\begin{align}
\operatorname{Var}(\mu_m)&=\frac{\mu_{2,m}}{n},\\
\operatorname{Var}(V_m)&=\frac{\mu_{4,m}-\mu_{2,m}^2}{n},\\
\operatorname{Cov}(\mu_m,V_m)&=\frac{\mu_{3,m}}{n}.
\end{align}
Accordingly,
\begin{align}
{\rm SEM}(\mu_m)&=\frac{\sigma_m}{\sqrt n},\\
{\rm SEM}(\sigma_m)&=\frac{\sqrt{\operatorname{Var}(V_m)}}{2\sigma_m}.
\end{align}
For $g(\mu,V)$, the delta-method variance is
\begin{equation}
\operatorname{Var}(g)\simeq
(\nabla g)^T
\begin{pmatrix}
\operatorname{Var}(\mu)&\operatorname{Cov}(\mu,V)\\
\operatorname{Cov}(\mu,V)&\operatorname{Var}(V)
\end{pmatrix}
\nabla g.
\end{equation}
The gradients used are
\begin{align}
\nabla C&=\left(-\frac{\sigma}{\mu^2},\frac{1}{2\mu\sigma}\right),\\
\nabla F&=\left(-\frac{V}{\mu^2},\frac{1}{\mu}\right).
\end{align}
Ratio and difference errors combine ON and OFF estimates under an independent-sample assumption. Distinct fixed seeds are verified for corrected CP5; the complete historical seed histories are not available for every baseline or Monash/A14 segment. These error bars quantify finite Monte Carlo precision only. They exclude tune,
generator-model, selection, particle-definition, fit-range, and detector
uncertainties.

\section{Association and fitted response}
Pearson's coefficient is
\begin{equation}
r=\frac{\sum_i(m_i-\bar m)(N_{{\rm ch},i}-\overline N_{\rm ch})}
{\sqrt{\sum_i(m_i-\bar m)^2\sum_i(N_{{\rm ch},i}-\overline N_{\rm ch})^2}},
\end{equation}
and measures linear association. Spearman's $\rho$ is Pearson's coefficient
applied to the ranks and measures monotonic ordering. The event-level fitted
response is $\nch=a+b\nmpi$, where $a$ is the intercept and $b$ is the
least-squares change in fitted multiplicity for a one-unit increase in
$\nmpi$. For fits to conditional profile points, the code uses weights
$\sqrt{N_m}$, corresponding to greater influence for more populated bins.
None of these quantities establishes causality, physical superiority, or a
one-to-one conversion of an additional interaction into final particles.

\begin{table}[htbp]\centering\small
\caption{Baseline association and slope summaries for INEL$>0$-like events with $n_{\rm MPI}>2$. Pearson's $r$, Spearman's $\rho$, and the event-level fitted slope $b$ are defined in this section. The global width summaries are retained in the Principal numerical summaries section.}
\begin{tabular}{llrrr}\toprule
$\sqrt{s}$&CR&$r$&$\rho$&$b$\\\midrule
2.76 TeV&ON&0.635&0.616&1.533\\
2.76 TeV&OFF&0.789&0.758&2.700\\
7 TeV&ON&0.714&0.703&1.697\\
7 TeV&OFF&0.850&0.830&3.078\\
13 TeV&ON&0.757&0.750&1.792\\
13 TeV&OFF&0.880&0.864&3.301\\
\bottomrule\end{tabular}\end{table}

\section{Measurable activity estimators}
The five final-state estimators are:
\begin{enumerate}
\item midrapidity multiplicity: charged particles with $|\eta|<0.8$ and
$\pt>0.15$ GeV/$c$;
\item forward multiplicity: the sum in $2.5<\eta<5.0$ and
$-5.0<\eta<-2.5$, with $\pt>0.15$ GeV/$c$;
\item transverse multiplicity: accepted central charged particles satisfying
$\pi/3<|\Delta\phi|<2\pi/3$ relative to the leading charged particle;
\item transverse scalar-$\pt$ sum: the scalar $\pt$ sum of the same transverse
particles;
\item forward--backward asymmetry:
$A_{\rm FB}=|N_F-N_B|/(N_F+N_B)$, evaluated only when $N_F+N_B>0$.
\end{enumerate}
The baseline five-estimator sample requires $\nmpi>2$ and positive midrapidity multiplicity without an upper momentum cut. The transverse observables additionally require a leading particle. The separate tune-estimator sample instead uses positive central multiplicity with $0.15<\pt<20$ GeV/$c$ as its event requirement; the estimators themselves retain no upper momentum cut. These separate generated samples and event definitions account for the nonidentical Monash entries in the two correlation tables.

The leading particle satisfies $|\eta|<0.8$ and $\pt>0.5$ GeV/$c$. Activity
classes are low, middle, and high quantiles of the chosen estimator. They enrich
different $\nmpi$ populations statistically; they do not measure $\nmpi$ event
by event. All estimators also receive contributions from ISR, FSR, the hard
process, beam remnants, hadronization, decays, and acceptance.

\section{Numerical values for the activity-class figure}
Table~\ref{tab:activity-classes} gives the numerical values plotted in the
manuscript activity-class figure. The low, middle, and high classes are formed
from quantiles of each estimator separately for CR ON and CR OFF. Their
estimator boundaries therefore need not be identical between the two CR
configurations. For integer multiplicities, the lower class includes the lower
boundary, the middle class is above the first boundary and includes the second,
and the high class is above the second boundary. The scalar-$\pt$ boundaries
and means are in GeV/$c$.

\begin{table}[htbp]
\centering
\small
\caption{Mean PYTHIA $\nmpi$ in the 13-TeV activity classes used in the
manuscript figure. $N$ is the number of events in the class, and the quoted
uncertainty is ${\rm SEM}=\sigma(\nmpi)/\sqrt{N}$.}
\label{tab:activity-classes}
\begin{tabular}{llcrrr}
\toprule
Estimator & CR & Class & Range & $N$ & $\langle\nmpi\rangle\pm{\rm SEM}$ \\
\midrule
Forward $N_{\rm ch}$ & ON  & low  & 0--32   & 527361 & $4.1396\pm0.0020$ \\
                     & ON  & mid  & 32--53  & 529846 & $6.8247\pm0.0036$ \\
                     & ON  & high & 53--192 & 507093 & $12.1790\pm0.0055$ \\
                     & OFF & low  & 0--40   & 524345 & $3.8754\pm0.0015$ \\
                     & OFF & mid  & 40--73  & 530686 & $6.6438\pm0.0028$ \\
                     & OFF & high & 73--273 & 511580 & $12.5562\pm0.0050$ \\
\addlinespace
Transverse $N_{\rm ch}$ & ON  & low  & 0--3   & 634134 & $5.2095\pm0.0032$ \\
                        & ON  & mid  & 3--6   & 460633 & $7.5865\pm0.0053$ \\
                        & ON  & high & 6--37  & 435504 & $11.5976\pm0.0067$ \\
                        & OFF & low  & 0--4   & 550332 & $4.4474\pm0.0023$ \\
                        & OFF & mid  & 4--9   & 524454 & $7.0290\pm0.0039$ \\
                        & OFF & high & 9--103 & 468855 & $12.3012\pm0.0058$ \\
\addlinespace
Transverse $\sum\pt$ & ON  & low  & 0--1.308     & 510092 & $4.9398\pm0.0032$ \\
                      & ON  & mid  & 1.308--3.237 & 510089 & $6.9961\pm0.0047$ \\
                      & ON  & high & 3.237--51.167& 510090 & $11.2931\pm0.0062$ \\
                      & OFF & low  & 0--1.858     & 514547 & $4.4420\pm0.0024$ \\
                      & OFF & mid  & 1.858--4.322 & 514547 & $6.8592\pm0.0040$ \\
                      & OFF & high & 4.322--73.432& 514547 & $11.8286\pm0.0058$ \\
\bottomrule
\end{tabular}
\end{table}

\section{Analysis workflow}
The calculation proceeds in the following order:
\begin{enumerate}
\item load the event metadata, accepted $\nch$ array, and INEL$>0$-like mask;
\item apply the baseline mask INEL$>0$-like and $\nmpi>2$ for the generator
diagnostic;
\item group events by integer $m=\nmpi$;
\item retain groups with $N_m\ge500$ for profile display only;
\item calculate $\mu_m$, central moments through fourth order, $\sigma_m$,
$C_m$, $F_m$, and their statistical standard errors;
\item calculate event-level global width, mean, correlations, and slope using all selected events, including bins below the display threshold;
\item merge CR-ON and CR-OFF profiles on the same integer $m$;
\item calculate matched-bin ratios and their common-bin weighted averages;
\item repeat normalized profiles for Monash, A14, and CMS CP5;
\item repeat global and fitted quantities for the selection and fit-range
sensitivity variants;
\item export CSV tables and construct the manuscript figures.
\end{enumerate}

\section{Principal numerical summaries}
\begin{table}[ht]
\centering\small
\caption{Global baseline results for INEL$>0$-like events with $\nmpi>2$.}
\begin{tabular}{llrrrrrr}\toprule
$\sqrt{s}$&CR&$N$&$\langle\nch\rangle$&$\sigma_{\rm ridge}$&$C_{\rm glob}$&$F_{\rm glob}$&$r$\\\midrule
2.76 TeV&ON&1311399&11.6469&5.0568&0.4342&2.1956&0.6354\\
2.76 TeV&OFF&1310744&15.8770&5.7497&0.3621&2.0822&0.7887\\
7 TeV&ON&1468690&14.5105&5.9966&0.4133&2.4782&0.7138\\
7 TeV&OFF&1469767&20.6824&6.8681&0.3321&2.2807&0.8504\\
13 TeV&ON&1566119&16.8036&6.7264&0.4003&2.6925&0.7569\\
13 TeV&OFF&1567470&24.6117&7.7426&0.3146&2.4357&0.8801\\\bottomrule
\end{tabular}
\end{table}

\begin{table}[ht]
\centering\small
\caption{Matched-bin ratios. The final column counts bins in which the CR-ON coefficient of variation exceeds the CR-OFF value.}
\begin{tabular}{lrrrrrr}\toprule
$\sqrt{s}$&range&bins&$\overline R_\sigma$&$\overline R_C$&$\overline R_F$&$C_{\rm ON}>C_{\rm OFF}$\\\midrule
2.76 TeV&3--18&16&0.8948&1.1739&1.0482&16/16\\
7 TeV&3--22&20&0.8912&1.2105&1.0760&20/20\\
13 TeV&3--25&23&0.8896&1.2322&1.0928&23/23\\\bottomrule
\end{tabular}
\end{table}

At 13 TeV, CR ON lowers the global conditional multiplicity scale from 24.6117
to 16.8036 (31.7\%) and the absolute ridge width from 7.7426 to 6.7264
(13.1\%). Since the mean decreases much more strongly, the global width-to-mean
ratio increases from 0.3146 to 0.4003. The matched-bin result verifies that this
reversal persists across the populated $\nmpi$ range and is not produced by a
single aggregate.

\section{Tune summary}
\begin{table}[ht]
\centering\small\color{black}
\caption{Event-weighted normalized fluctuation measures at 13 TeV for retained
bins with $\nmpi>2$.}
\begin{tabular}{llrrr}\toprule
Tune&CR&$\nmpi$ range&event-weighted $C_m$&event-weighted $F_m$\\\midrule
Monash &ON&3--25&0.4204&2.8083\\
Monash &OFF&3--25&0.3495&2.5755\\
ATLAS A14 &ON&3--26&0.4139&2.6748\\
ATLAS A14 &OFF&3--26&0.3385&2.4155\\
{\color{black} CMS CP5} &ON&3--44&0.4363&2.5954\\
{\color{black} CMS CP5} &OFF&3--44&0.3580&2.5974\\\bottomrule
\end{tabular}
\end{table}

\section{Interpretation and limitations}
The calculation supports the following precise statement: within the specified
PYTHIA configurations and at the same integer $\nmpi$, CR ON produces a lower
conditional mean and a smaller absolute conditional width, but the width falls
less strongly than the mean; hence relative dispersion is larger with CR ON.
The result is stable across the tested energies and tunes, while its numerical
magnitude remains model dependent.

The analysis does not establish that CR ON or CR OFF is globally superior, does
not measure MPI experimentally, does not isolate the direct causal action of CR,
and does not reproduce a detector-level ALICE selection. A dedicated CR-OFF
retune, provenance-complete primary-particle samples, detector response, other
generator models, and broader systematic variations would be required to extend
those conclusions.

\section{Inclusive distributions and estimator-profile inputs}
For event $i$ with $N_{{\rm ch},i}>0$, the event-mean transverse momentum is
\begin{equation}
\langle p_{\rm T}\rangle_{{\rm event},i}
=\frac{1}{N_{{\rm ch},i}}\sum_{j=1}^{N_{{\rm ch},i}}p_{{\rm T},ij}.
\end{equation}
Here $p_{{\rm T},ij}$ is the transverse momentum of the $j$th accepted charged particle.

The inclusive distributions in manuscript Figs.~1 and 2 use the INEL$>0$-like sample without a selection on $\nmpi$. The multiplicity distribution retains zero central multiplicity. The event-mean-$\pt$ distribution additionally requires $\nch>0$ and assigns equal weight to each event. Its bin width is 0.5 GeV/$c$, while the multiplicity-bin width is 5. The machine-readable histograms and sample summaries are supplied in Supplementary Material S2.

The five-estimator maps (manuscript Fig.~7) retain all selected events and use raw events per bin with logarithmic color normalization. Conditional means are displayed for $N_m\ge500$. Pearson coefficients and ordinary least-squares fits use all selected events, rather than the conditional profile points or histogram-bin centers.

For the tune profiles (manuscript Fig.~9), the standard error is $s_m/\sqrt{N_m}$, where $s_m^2=\sum_i(x_i-\bar x_m)^2/(N_m-1)$ for the estimator $x$. This finite-sample convention differs from the population-moment convention used for the main conditional-width and activity-class calculations by a factor $\sqrt{N_m/(N_m-1)}$, below 0.11\% for displayed bins. The error bars quantify uncertainty in the mean, not the event spread. {\color{black} The twelve Monash/A14 estimator coefficients reproduce their archived values to $10^{-10}$. The six CP5 coefficients and their profiles are recalculated from the freshly generated CP5 events; the obsolete approximate-CP5 coefficients are excluded.}

\section{Matched-bin verification across tunes}
At 13 TeV, the coefficient of variation is larger with CR ON at every common retained integer $\nmpi$ in each tested tune. Common-bin weights are $w_m=\min(N_{m,{\rm ON}},N_{m,{\rm OFF}})$, as in the energy comparison.
\begin{table}[htbp]\centering
\color{black}\caption{Matched-bin checks in the independent tune samples. The last column counts bins with $C_{m,{\rm ON}}>C_{m,{\rm OFF}}$.}
\begin{tabular}{lrrrr}\toprule
Tune & $\nmpi$ range & $\overline R_\sigma$ & $\overline R_C$ & Passing bins\\\midrule
Monash & 3--25 & 0.8902 & 1.2275 & 23/23\\
ATLAS A14 & 3--26 & 0.8878 & 1.2503 & 24/24\\
{\color{black} CMS CP5 } & 3--44 & 0.8002 & 1.2557 & 42/42\\\bottomrule
\end{tabular}\end{table}
The corresponding CSV is \texttt{tune\_matched\_bin\_verification.csv} in Supplementary Material S2. These ratios differ from averages of $C_m$ evaluated separately in each CR sample.

\appendix
\begin{landscape}
\section{Complete conditional profiles}
The following table reproduces every row of
\texttt{fresh\_conditional\_fluctuations\_by\_nmpi.csv}. ``SE'' denotes the
one-standard-error Monte Carlo uncertainty.
\scriptsize
\begin{longtable}{llrrrrrrrrrr}
\toprule
Energy&CR&$m$&$N_m$&$\mu_m$&$\sigma_m$&$C_m$&$F_m$&SE($\mu$)&SE($\sigma$)&SE($C$)&SE($F$)\\\midrule
\endfirsthead
\toprule Energy&CR&$m$&$N_m$&$\mu_m$&$\sigma_m$&$C_m$&$F_m$&SE($\mu$)&SE($\sigma$)&SE($C$)&SE($F$)\\\midrule
\endhead
13 TeV&CR OFF&3&275123&\num{9.53729422839966}&\num{4.682495536489548}&\num{0.4909668742887532}&\num{2.298950197421312}&\num{0.0089271724195154}&\num{0.0078291385148636}&\num{0.0007492022303071}&\num{0.0070380234821786}\\
13 TeV&CR OFF&4&207381&\num{12.642995259932205}&\num{5.481328722850128}&\num{0.4335466881191823}&\num{2.37641191428422}&\num{0.0120365317546904}&\num{0.0111208814171542}&\num{0.0008239970552693}&\num{0.0090649320747485}\\
13 TeV&CR OFF&5&166560&\num{15.789199087415946}&\num{6.226064430847122}&\num{0.3943242716984498}&\num{2.455088322241415}&\num{0.0152555635229294}&\num{0.0429405106065226}&\num{0.0026650506377887}&\num{0.0334429620952145}\\
13 TeV&CR OFF&6&140298&\num{18.99018517726553}&\num{6.788819191097694}&\num{0.3574909421749641}&\num{2.4269413688809927}&\num{0.0181246029792843}&\num{0.0148504537340452}&\num{0.0007420860461709}&\num{0.0100878237219717}\\
13 TeV&CR OFF&7&119757&\num{22.26945397763805}&\num{7.403535973556347}&\num{0.3324525145964797}&\num{2.4613241513143045}&\num{0.0213938396762427}&\num{0.0183326799960498}&\num{0.0007841776038074}&\num{0.011666737721995}\\
13 TeV&CR OFF&8&105923&\num{25.55275058297065}&\num{7.940869579932925}&\num{0.3107637885850547}&\num{2.467734715319768}&\num{0.0243990505610906}&\num{0.0202668523252135}&\num{0.0007627265992678}&\num{0.0121305417624286}\\
13 TeV&CR OFF&9&93574&\num{28.854468121486736}&\num{8.429004284291036}&\num{0.2921212842601074}&\num{2.4622915565610457}&\num{0.0275548897221283}&\num{0.0215668833580699}&\num{0.0007174879196096}&\num{0.0121245283429694}\\
13 TeV&CR OFF&10&83073&\num{32.240378943820495}&\num{8.95022453576209}&\num{0.2776091605920028}&\num{2.4846643204828616}&\num{0.0310530300446712}&\num{0.0239275170009875}&\num{0.0007182226136696}&\num{0.012851584990199}\\
\end{longtable}
\end{landscape}

\begin{landscape}
\section{Complete selection and fit-range sensitivity table}
\begin{table}[!tbp]\centering\small
\caption{\label{tab:fit_sensitivity}13~TeV Monash slope sensitivity. Event-level fits and conditional-profile fits are different estimators and are compared only within the same row pair.}
\begin{tabular}{lrr}\toprule
Selection / fit&CR ON&CR OFF\\\midrule
Event fit: all selected $n_{\rm MPI}$&1.858&3.168\\
Event fit: $n_{\rm MPI}>2$&1.792&3.301\\
Relative change [\%]&$-3.5$&$+4.2$\\\midrule
Profile fit: retained bins&1.794&3.301\\
Profile fit: up to 90th percentile&1.859&3.280\\
Change magnitude [\%]&3.6&0.6\\\bottomrule
\end{tabular}\end{table}

Dashes indicate quantities not applicable to profile-only fits.
\scriptsize
\begin{longtable}{rlrrrrrp{6.8cm}}
\toprule
$N$&Energy / CR&mean&width&$C_{\rm glob}$&$r$&slope&selection\\\midrule
\endfirsthead
\toprule $N$&Energy / CR&mean&width&$C_{\rm glob}$&$r$&slope&selection\\\midrule\endhead
3309568&2.76 TeV\ / CR ON&\num{6.922446071511448}&\num{3.7918351113112}&\num{0.5477594295629218}&\num{0.7754175382434552}&\num{1.6567842932069552}&INEL>0; all nMPI\\
1782313&2.76 TeV\ / CR ON&\num{9.951714429508172}&\num{4.630150469192603}&\num{0.4652615890447564}&\num{0.7002667434869065}&\num{1.5900093088311322}&INEL>0; nMPI>1\\
1311399&2.76 TeV\ / CR ON&\num{11.646861862789281}&\num{5.0568214477482}&\num{0.43417888074248634}&\num{0.6354461105563975}&\num{1.5325900913178065}&INEL>0; nMPI>2\\
1308820&2.76 TeV\ / CR ON&\num{11.669811738818172}&\num{5.049341611190322}&\num{0.43268406759248035}&\num{0.6349079110778167}&\num{1.5277821975868895}&Nch>0 and nMPI>2\\
1310994&2.76 TeV\ / CR ON&---&---&---&---&\num{1.5342253882799475}&profile bins: all with >=500 events\\
1221266&2.76 TeV\ / CR ON&---&---&---&---&\num{1.5979082128108453}&profile bins: nMPI<=90th percentile\\
1302244&2.76 TeV\ / CR ON&---&---&---&---&\num{1.5477190419083757}&profile bins: nMPI<=99th percentile\\
3307698&2.76 TeV\ / CR OFF&\num{8.740807655354267}&\num{4.2626278417954895}&\num{0.487669790924226}&\num{0.8678378525195823}&\num{2.634078817480701}&INEL>0; all nMPI\\
\end{longtable}
\end{landscape}

\begin{landscape}
\section{Complete tune-normalized profiles}\color{black}
This table reproduces every row of
\texttt{fresh\_tune\_normalized\_fluctuation\_profiles.csv}, with at least two events per conditional bin. Singleton bins remain in the comprehensive machine-readable results but are excluded here because they cannot constrain a width uncertainty. The manuscript summary applies $N_m\ge500$ and $m>2$.
\scriptsize
\begin{longtable}{rrrrlrr}
\toprule
$m$&$\mu_m$&$\sigma_m$&$N_m$&sample&$C_m$&$F_m$\\\midrule
\endfirsthead
\toprule $m$&$\mu_m$&$\sigma_m$&$N_m$&sample&$C_m$&$F_m$\\\midrule\endhead
0&\num{0.3636363636363636}&\num{0.4810456929208346}&22&ATLAS A14 CR OFF&\num{1.3228756555322954}&\num{0.6363636363636364}\\
1&\num{4.741402014454393}&\num{5.007605996584172}&1362354&ATLAS A14 CR OFF&\num{1.0561445710189188}&\num{5.288755887094156}\\
2&\num{6.751562480351102}&\num{3.760263166892696}&397605&ATLAS A14 CR OFF&\num{0.5569471034054847}&\num{2.0942676788432224}\\
3&\num{9.911513029224952}&\num{4.784502104398707}&268435&ATLAS A14 CR OFF&\num{0.482721668255007}&\num{2.3095828376049354}\\
4&\num{13.147055619919}&\num{5.597294876866248}&201978&ATLAS A14 CR OFF&\num{0.4257451279346403}&\num{2.383021023439328}\\
5&\num{16.468509918499155}&\num{6.344564493331523}&162575&ATLAS A14 CR OFF&\num{0.3852543141261763}&\num{2.444270842307727}\\
6&\num{19.853989417835383}&\num{6.938471842473665}&137401&ATLAS A14 CR OFF&\num{0.3494749441258715}&\num{2.424822059467416}\\
7&\num{23.3468961040056}&\num{7.574321816736424}&118609&ATLAS A14 CR OFF&\num{0.3244252162255053}&\num{2.4573009931562764}\\
\end{longtable}
\end{landscape}

\section{Files and provenance}
The principal calculation is archived at
\path{submission_tools/reproduce_conditional_results.py}.
{\color{black} Current generation sources are in \path{generation/}; original CP5 workflow programs are retained as historical provenance in \path{archived_analysis_code/cp5_historical_workflow/}. Production-time settings and particle data are in \path{cp5_exact/production_settings/}; sixteen independent seeds and available logs are in \path{production_records/}. Complete event-summary files remain externally retained and identified by source hashes; lossless joint-count inputs for the conditional analysis are supplied in \path{reproduction_data/}. Both tune analyses use the same fresh CP5 samples. The external PDF is NNPDF31\_nnlo\_as\_0118 member 0, accessed with LHAPDF 6.5.4 in PYTHIA 8.316.}\par The exported numerical files are in
\texttt{new\_analysis\_editor\_response/}. The archive also contains historical acceptance, ALICE comparison, activity-estimator, tune-generation, tune-analysis, and output-assembly programs. Historical workflows are provenance records rather than supported execution instructions; current portable commands are listed in the S2 README. Large generated event arrays are available from
the corresponding author upon reasonable request and are not embedded in this
PDF. Supplementary Material S2 also includes the inclusive and estimator plotting programs, their numerical inputs, and a script for reproducing the revised estimator figures without the large event samples. The Supplementary Material S1 and manuscript use identical definitions; the explicit primary-particle limitation remains unresolved by the available records.

\clearpage
\section{Complete tune settings and production details}
\begingroup\color{black}
The manuscript retains the key tune settings and a short explanation of their physical roles. This section records the complete comparison table and implementation details moved out of the main prose. They support reproducibility without changing the analysis.

\begin{table}[!tbp]\centering\color{black}\footnotesize
\caption{\label{tab:complete_settings}Settings of the independently generated 13~TeV configuration scan in PYTHIA 8.316. Monash and A14 values are reconstructed from archived source; CP5 values are saved during the fresh production. All configurations use \texttt{SoftQCD:all = on}; only the physics setting \texttt{ColourReconnection:reconnect} changes between ON and OFF; distinct fixed seeds are documented for the initial Monash/A14 commands and all corrected CP5 segments; complete historical top-up seed linkage is not certified. The baseline is associated with a Monash source family, but no complete per-run production configuration is certified. Candidate sources include lifetime and species decay overrides, and actual historical seeds are unavailable in retained records; these candidates must not be presented as verified effective settings.}
\resizebox{\textwidth}{!}{%
\begin{tabular}{lccc}\toprule
Setting & Monash & A14 & CMS CP5\\\midrule
\texttt{Tune:pp} (starting point)&14&21&14 + overrides\\
\texttt{PDF:pSet}&13&13&LHAPDF6 (member 0)\\
PDF family &NNPDF2.3 LO QCD+QED&NNPDF2.3 LO QCD+QED&NNPDF3.1 NNLO\\
MPI $p_{\rm T0}^{\rm ref}$ [GeV]&2.28&2.09&1.41\\
MPI reference energy [GeV]&7000&7000&7000\\
MPI energy exponent&0.215&0.215&0.03344\\
\texttt{MultipartonInteractions:bProfile}&3&3&2\\
Overlap exponent (profile 3)&1.85&1.85&--\\
Core radius / fraction (profile 2)&--&--&0.7634 / 0.63\\
\texttt{ColourReconnection:mode}&0&0&0\\
\texttt{ColourReconnection:range}&1.80&1.71&5.176\\
$\alpha_s(M_Z)$: hard process / running order&0.130 / 1&0.140 / 1&0.118 / 2\\
$\alpha_s(M_Z)$: ISR / running order&0.1365 / 1&0.127 / 1&0.118 / 2\\
$\alpha_s(M_Z)$: FSR / running order&0.1365 / 1&0.127 / 1&0.118 / 2\\
$\alpha_s(M_Z)$: MPI / running order&0.130 / 1&0.126 / 1&0.118 / 2\\
\texttt{SpaceShower:rapidityOrder}&on&on&on\\
\texttt{SigmaDiffractive:mode}&1&1&1\\
\texttt{SigmaTotal:zeroAXB}&on&on&off\\
\texttt{SigmaTotal:mode}&1&1&0\\
\texttt{SigmaTotal:sigmaTot} [mb]&100&100&100.309\\
\texttt{SigmaTotal:sigmaEl} [mb]&25&25&21.89\\\bottomrule
\end{tabular}}

\end{table}

The CP5 PDF is \texttt{LHAPDF6:NNPDF31\_nnlo\_as\_0118}, member 0, accessed with LHAPDF 6.5.4. The four coupling sectors use two-loop running with $\alpha_s(M_Z)=0.118$. Central diffraction is retained with \texttt{SigmaTotal:zeroAXB = off}. The full-precision parameters are $p_{\rm T0}^{\rm ref}=1.41$ GeV, energy exponent 0.03344, core radius 0.7634, core fraction 0.63 and CR range 5.176. The total-cross-section settings are \texttt{SigmaTotal:mode = 0}, $\sigma_{\rm tot}=100.309$ mb and $\sigma_{\rm el}=21.89$ mb. The published settings are implemented in PYTHIA 8.316; the original CMS PYTHIA 8.226 environment is not reproduced.

The MPI regularization and energy scaling follow the PYTHIA 8.3 guide by Bierlich et al. (cited in the main manuscript):
\begin{equation}
\frac{\mathrm{d}\sigma}{\mathrm{d}p_{\rm T}^2}\propto
\frac{\alpha_s^2(p_{\rm T}^2+p_{\rm T0}^2)}{(p_{\rm T}^2+p_{\rm T0}^2)^2},\qquad
p_{\rm T0}=p_{\rm T0}^{\rm ref}\left(\frac{\sqrt{s}}{E_{\rm ref}}\right)^{\epsilon}.
\end{equation}
The reference scale regulates the low-$p_{\rm T}$ interaction rate, while $\epsilon$ controls its energy dependence.

For profile 3, the overlap has the form $\mathcal{O}(b)\propto\exp[-(b/b_0)^p]$, with $p=\texttt{expPow}$. Profile 2 is a double-Gaussian matter distribution whose narrow component is set by \texttt{coreRadius} and \texttt{coreFraction}. In CR mode 0, a softer system of scale $p_{\rm T}$ reconnects with probability $p_{\rm T,rec}^2/(p_{\rm T,rec}^2+p_{\rm T}^2)$, where $p_{\rm T,rec}=R p_{\rm T0}$ and $R=\texttt{ColourReconnection:range}$. The $m_0$ and time-dilation parameters of mode 1 do not enter these mode-0 comparisons.

The CP5 pair uses the same default PYTHIA decay prescription without particle-specific overrides. Eight independent fixed seeds contribute five million events to each CR state: four verified logged prefixes from an interrupted production plus four new independent-seed runs with unchanged physics settings. Recovered prefixes lack final end-of-run warning summaries; checkpoint validation and complete top-up logs are retained. Per-run counts, seeds, settings, particle data and source records are supplied with the analysis material. The same CP5 events support both fluctuation and estimator analyses. Monash and A14 settings are reconstructed from archived source; the historical six-baseline decay settings and seeds remain subject to production-record confirmation.
\endgroup

\section{Production provenance and independent reproducibility}
The twelve central analysis samples are distinguished in S2's
\path{production_records/sample_provenance_matrix.csv}. All six original
baseline samples retain five million metadata records and their particle
chunks. Candidate generation sources include
\texttt{ParticleDecays:limitTau0 = on}, \texttt{tau0Max = 10.0}, and decay
overrides for species 333, 310, 3122, 3312 and 3334 and the specified
antiparticles. The source-to-run link, production executable, complete effective
settings, actual time-derived seeds and full historical logs have not been
recovered as auditable per-run records. These candidates are supplied as
unverified evidence, rather than substituted for missing production records.
Thus exact historical generator replay is not established. The retained-event
selection and conditional numerical analysis can nevertheless be independently
checked, and the supplied joint counts reproduce the conditional moments and
uncertainties without event-order information. This distinction does not
establish that a newly generated default-decay sample is equivalent to the
historical baseline.

Monash and A14 tune configurations are reconstructed from archived source;
available shell commands and logs are supplied, but complete seed-to-segment
linkage for all historical top-ups is not certified. Corrected CP5 has complete
per-segment settings and particle dumps, sixteen fixed seeds, checked event-file
hashes and available checkpoint/final logs. Interrupted prefixes lack final
warning summaries and are identified as such. The retained CMS card, PDF
metadata and central PDF grid accompany the production records. Fresh short
generator validation runs are kept only in the author check directory; they
are not production samples and do not establish missing historical provenance.

To reproduce the conditional results from the supplied integer joint counts,
run \path{python submission_tools/reproduce_conditional_results.py} from S2.
The outputs include central moments through fourth order, errors for every
conditional mean/width/coefficient of variation/scaled variance, and every
retained ON/OFF matched-bin comparison. Histogram aggregation preserves all
information needed for these one-event conditional statistics, but does not
preserve particle ancestry, generator replay, event ordering or time correlations.

\section{Every retained matched bin: ratios and statistical uncertainties}
The following tables use the common population-moment convention and empirical moments through fourth order. The quoted errors are marginal one-standard-error delta-method estimates; CR samples are treated as independent. $N_+$ and $N_-$ denote ON and OFF counts. All 148 central-value orderings satisfy $R_\mu<R_\sigma<1$, hence $R_C>1$. The two signed differences, $\sigma_- - \sigma_+$ and $C_+ - C_-$, also pass a Bonferroni normal-approximation diagnostic across the 296 directional comparisons, with family error target 0.05. This is an asymptotic diagnostic, not an exact finite-sample guarantee. The machine-readable file additionally supplies conditional means, widths, coefficients of variation, their errors, ratio confidence intervals and central moments through fourth order.
\subsection*{Baseline: 13 TeV}
\begingroup\scriptsize\setlength{\tabcolsep}{3pt}
\begin{longtable}{rrrrrrrrr}
\toprule $m$ & $N_+$ & $N_-$ & $R_\mu$ & SE & $R_\sigma$ & SE & $R_C$ & SE\\\midrule\endhead
3 & 274585 & 275123 & 0.846281 & 0.001186 & 0.942415 & 0.002311 & 1.113596 & 0.002428\\
4 & 207204 & 207381 & 0.801095 & 0.001165 & 0.924434 & 0.002651 & 1.153963 & 0.003006\\
5 & 166510 & 166560 & 0.765894 & 0.001142 & 0.900605 & 0.006515 & 1.175887 & 0.008271\\
6 & 139540 & 140298 & 0.736594 & 0.001111 & 0.899477 & 0.002877 & 1.221130 & 0.003604\\
7 & 120063 & 119757 & 0.713076 & 0.001092 & 0.885776 & 0.003099 & 1.242191 & 0.004038\\
8 & 105978 & 105923 & 0.692569 & 0.001068 & 0.878552 & 0.003208 & 1.268541 & 0.004333\\
9 & 93432 & 93574 & 0.679153 & 0.001054 & 0.869323 & 0.003271 & 1.280010 & 0.004516\\
10 & 82993 & 83073 & 0.663334 & 0.001048 & 0.862401 & 0.003371 & 1.300100 & 0.004795\\
11 & 73050 & 73396 & 0.649665 & 0.001048 & 0.855042 & 0.003557 & 1.316127 & 0.005172\\
12 & 64080 & 64169 & 0.638057 & 0.001053 & 0.847072 & 0.003745 & 1.327580 & 0.005580\\
13 & 55221 & 54977 & 0.629479 & 0.001081 & 0.848811 & 0.003951 & 1.348433 & 0.005991\\
14 & 45941 & 45973 & 0.618663 & 0.001123 & 0.839172 & 0.004285 & 1.356427 & 0.006602\\
15 & 37186 & 37389 & 0.611226 & 0.001195 & 0.838126 & 0.004782 & 1.371222 & 0.007479\\
16 & 29513 & 29283 & 0.601946 & 0.001280 & 0.822711 & 0.005340 & 1.366753 & 0.008525\\
17 & 22215 & 22429 & 0.595581 & 0.001421 & 0.821777 & 0.005994 & 1.379790 & 0.009663\\
18 & 16355 & 16163 & 0.586786 & 0.001589 & 0.809200 & 0.006739 & 1.379038 & 0.011070\\
19 & 11620 & 11396 & 0.581606 & 0.001826 & 0.814684 & 0.008167 & 1.400748 & 0.013478\\
20 & 7767 & 7789 & 0.570228 & 0.002144 & 0.799838 & 0.009699 & 1.402663 & 0.016354\\
21 & 5164 & 5168 & 0.564242 & 0.002535 & 0.801273 & 0.012007 & 1.420088 & 0.020594\\
22 & 3118 & 3213 & 0.559259 & 0.003155 & 0.783353 & 0.014519 & 1.400698 & 0.025011\\
23 & 2035 & 1995 & 0.556214 & 0.003843 & 0.798615 & 0.018541 & 1.435806 & 0.032189\\
24 & 1143 & 1144 & 0.550160 & 0.004849 & 0.789847 & 0.024390 & 1.435668 & 0.043155\\
25 & 646 & 610 & 0.554223 & 0.006584 & 0.751916 & 0.032176 & 1.356705 & 0.056256\\
\bottomrule\end{longtable}\endgroup
\subsection*{Baseline: 2.76 TeV}
\begingroup\scriptsize\setlength{\tabcolsep}{3pt}
\begin{longtable}{rrrrrrrrr}
\toprule $m$ & $N_+$ & $N_-$ & $R_\mu$ & SE & $R_\sigma$ & SE & $R_C$ & SE\\\midrule\endhead
3 & 315094 & 314941 & 0.854492 & 0.001102 & 0.935319 & 0.002190 & 1.094591 & 0.002335\\
4 & 235445 & 234495 & 0.808700 & 0.001084 & 0.911399 & 0.002364 & 1.126993 & 0.002704\\
5 & 185295 & 184654 & 0.772455 & 0.001076 & 0.900855 & 0.002449 & 1.166223 & 0.002938\\
6 & 149698 & 148992 & 0.742602 & 0.001068 & 0.884945 & 0.002616 & 1.191681 & 0.003294\\
7 & 119865 & 119478 & 0.717681 & 0.001083 & 0.874083 & 0.002897 & 1.217928 & 0.003791\\
8 & 94076 & 94318 & 0.697138 & 0.001119 & 0.864506 & 0.003130 & 1.240077 & 0.004258\\
9 & 70994 & 70899 & 0.677977 & 0.001197 & 0.853173 & 0.003972 & 1.258410 & 0.005594\\
10 & 50799 & 51876 & 0.663498 & 0.001321 & 0.849730 & 0.004197 & 1.280682 & 0.005989\\
11 & 35616 & 35412 & 0.647563 & 0.001484 & 0.836897 & 0.004852 & 1.292380 & 0.007154\\
12 & 23071 & 23333 & 0.639478 & 0.001758 & 0.836750 & 0.006118 & 1.308489 & 0.009152\\
13 & 14052 & 14441 & 0.631886 & 0.002140 & 0.833400 & 0.007555 & 1.318909 & 0.011477\\
14 & 8239 & 8395 & 0.621194 & 0.002672 & 0.812526 & 0.009517 & 1.308008 & 0.014841\\
15 & 4488 & 4785 & 0.612423 & 0.003386 & 0.811625 & 0.012951 & 1.325269 & 0.020390\\
16 & 2514 & 2495 & 0.598064 & 0.004348 & 0.790565 & 0.016593 & 1.321873 & 0.026778\\
17 & 1182 & 1213 & 0.593249 & 0.006091 & 0.806793 & 0.022864 & 1.359958 & 0.038136\\
18 & 566 & 559 & 0.583609 & 0.008700 & 0.757660 & 0.034694 & 1.298231 & 0.057917\\
\bottomrule\end{longtable}\endgroup
\subsection*{Baseline: 7 TeV}
\begingroup\scriptsize\setlength{\tabcolsep}{3pt}
\begin{longtable}{rrrrrrrrr}
\toprule $m$ & $N_+$ & $N_-$ & $R_\mu$ & SE & $R_\sigma$ & SE & $R_C$ & SE\\\midrule\endhead
3 & 289852 & 290478 & 0.847656 & 0.001146 & 0.936243 & 0.002326 & 1.104508 & 0.002471\\
4 & 217803 & 217508 & 0.803933 & 0.001129 & 0.918473 & 0.002443 & 1.142474 & 0.002765\\
5 & 174812 & 175081 & 0.764953 & 0.001099 & 0.903264 & 0.002620 & 1.180809 & 0.003148\\
6 & 145770 & 145296 & 0.737943 & 0.001081 & 0.890752 & 0.002834 & 1.207075 & 0.003567\\
7 & 123848 & 123925 & 0.713262 & 0.001063 & 0.879869 & 0.003406 & 1.233585 & 0.004520\\
8 & 106379 & 106758 & 0.694033 & 0.001057 & 0.878612 & 0.003055 & 1.265952 & 0.004138\\
9 & 91677 & 91638 & 0.677171 & 0.001056 & 0.866547 & 0.003246 & 1.279657 & 0.004520\\
10 & 77497 & 77768 & 0.662374 & 0.001071 & 0.860437 & 0.003428 & 1.299019 & 0.004900\\
11 & 64015 & 64298 & 0.650277 & 0.001106 & 0.849836 & 0.003929 & 1.306883 & 0.005757\\
12 & 51296 & 50816 & 0.636524 & 0.001165 & 0.843531 & 0.004058 & 1.325215 & 0.006094\\
13 & 39457 & 39468 & 0.628695 & 0.001264 & 0.840879 & 0.004600 & 1.337498 & 0.006992\\
14 & 29088 & 29413 & 0.613851 & 0.001389 & 0.816053 & 0.005210 & 1.329398 & 0.008154\\
15 & 20671 & 20823 & 0.608977 & 0.001578 & 0.828411 & 0.006176 & 1.360332 & 0.009722\\
16 & 13941 & 13960 & 0.596472 & 0.001850 & 0.813792 & 0.007469 & 1.364342 & 0.012009\\
17 & 9245 & 9305 & 0.590273 & 0.002163 & 0.797132 & 0.008759 & 1.350445 & 0.014387\\
18 & 5742 & 5734 & 0.582768 & 0.002647 & 0.801313 & 0.011588 & 1.375013 & 0.019153\\
19 & 3432 & 3459 & 0.576487 & 0.003326 & 0.811940 & 0.014606 & 1.408429 & 0.024410\\
20 & 2027 & 1933 & 0.573549 & 0.004216 & 0.786016 & 0.018264 & 1.370443 & 0.031419\\
21 & 1044 & 1035 & 0.556708 & 0.005527 & 0.774401 & 0.025937 & 1.391036 & 0.044277\\
22 & 573 & 565 & 0.562504 & 0.007133 & 0.784353 & 0.033715 & 1.394395 & 0.058748\\
\bottomrule\end{longtable}\endgroup
\subsection*{Tune: ATLAS A14}
\begingroup\scriptsize\setlength{\tabcolsep}{3pt}
\begin{longtable}{rrrrrrrrr}
\toprule $m$ & $N_+$ & $N_-$ & $R_\mu$ & SE & $R_\sigma$ & SE & $R_C$ & SE\\\midrule\endhead
3 & 267771 & 268435 & 0.842283 & 0.001176 & 0.938461 & 0.003239 & 1.114188 & 0.003591\\
4 & 202108 & 201978 & 0.799434 & 0.001154 & 0.919167 & 0.002535 & 1.149773 & 0.002890\\
5 & 162705 & 162575 & 0.763359 & 0.001129 & 0.902680 & 0.004318 & 1.182511 & 0.005390\\
6 & 137101 & 137401 & 0.729939 & 0.001090 & 0.895926 & 0.002883 & 1.227398 & 0.003664\\
7 & 118606 & 118609 & 0.706753 & 0.001068 & 0.886332 & 0.003000 & 1.254090 & 0.003964\\
8 & 104537 & 104075 & 0.685673 & 0.001041 & 0.871143 & 0.003543 & 1.270493 & 0.004893\\
9 & 92727 & 92729 & 0.669112 & 0.001034 & 0.869066 & 0.005643 & 1.298835 & 0.008139\\
10 & 83128 & 83047 & 0.655550 & 0.001014 & 0.864631 & 0.004417 & 1.318940 & 0.006473\\
11 & 75304 & 74497 & 0.642197 & 0.001006 & 0.869580 & 0.003590 & 1.354071 & 0.005303\\
12 & 66843 & 66661 & 0.633281 & 0.001008 & 0.862624 & 0.003681 & 1.362151 & 0.005538\\
13 & 58653 & 59199 & 0.621464 & 0.001015 & 0.854297 & 0.003918 & 1.374653 & 0.006006\\
14 & 50886 & 51527 & 0.612861 & 0.001037 & 0.853816 & 0.004148 & 1.393165 & 0.006467\\
15 & 43074 & 43068 & 0.603597 & 0.001078 & 0.830182 & 0.005042 & 1.375391 & 0.008049\\
16 & 35155 & 35555 & 0.595121 & 0.001134 & 0.830264 & 0.005468 & 1.395118 & 0.008870\\
17 & 28322 & 28682 & 0.589780 & 0.001222 & 0.826986 & 0.005265 & 1.402194 & 0.008580\\
18 & 21746 & 22079 & 0.580184 & 0.001334 & 0.830861 & 0.006031 & 1.432065 & 0.010004\\
19 & 16517 & 16656 & 0.576495 & 0.001473 & 0.811460 & 0.006756 & 1.407576 & 0.011359\\
20 & 11879 & 12014 & 0.568421 & 0.001681 & 0.822450 & 0.008100 & 1.446903 & 0.013783\\
21 & 8134 & 8124 & 0.560111 & 0.001962 & 0.838581 & 0.009972 & 1.497169 & 0.017199\\
22 & 5509 & 5674 & 0.553552 & 0.002265 & 0.797062 & 0.011477 & 1.439904 & 0.020014\\
23 & 3449 & 3670 & 0.553108 & 0.002855 & 0.812183 & 0.014428 & 1.468398 & 0.025375\\
24 & 2253 & 2294 & 0.540629 & 0.003365 & 0.797101 & 0.016870 & 1.474396 & 0.030410\\
25 & 1415 & 1352 & 0.547602 & 0.004271 & 0.831668 & 0.022768 & 1.518746 & 0.040871\\
26 & 832 & 738 & 0.524410 & 0.005116 & 0.723553 & 0.026867 & 1.379745 & 0.049986\\
\bottomrule\end{longtable}\endgroup
\subsection*{Tune: CMS CP5}
\begingroup\scriptsize\setlength{\tabcolsep}{3pt}
\begin{longtable}{rrrrrrrrr}
\toprule $m$ & $N_+$ & $N_-$ & $R_\mu$ & SE & $R_\sigma$ & SE & $R_C$ & SE\\\midrule\endhead
3 & 221434 & 221216 & 0.824373 & 0.001427 & 0.892877 & 0.002761 & 1.083098 & 0.002926\\
4 & 165316 & 164980 & 0.776643 & 0.001419 & 0.868585 & 0.002804 & 1.118384 & 0.003177\\
5 & 131559 & 131473 & 0.735155 & 0.001398 & 0.847436 & 0.002999 & 1.152730 & 0.003634\\
6 & 109656 & 109599 & 0.698946 & 0.001368 & 0.832992 & 0.003179 & 1.191783 & 0.004089\\
7 & 93807 & 93355 & 0.673417 & 0.001348 & 0.818282 & 0.003347 & 1.215121 & 0.004525\\
8 & 82318 & 82007 & 0.651567 & 0.001330 & 0.806547 & 0.003374 & 1.237857 & 0.004732\\
9 & 73114 & 73221 & 0.634561 & 0.001315 & 0.799898 & 0.003620 & 1.260554 & 0.005214\\
10 & 66808 & 66635 & 0.617283 & 0.001287 & 0.785732 & 0.004243 & 1.272887 & 0.006443\\
11 & 60409 & 60507 & 0.605156 & 0.001274 & 0.780998 & 0.003863 & 1.290572 & 0.005932\\
12 & 56014 & 55854 & 0.592590 & 0.001259 & 0.775451 & 0.003811 & 1.308579 & 0.005981\\
13 & 52413 & 52000 & 0.582983 & 0.001245 & 0.763939 & 0.011121 & 1.310397 & 0.018629\\
14 & 48787 & 48566 & 0.574776 & 0.001228 & 0.767835 & 0.004434 & 1.335887 & 0.007272\\
15 & 45808 & 45886 & 0.566551 & 0.001211 & 0.762605 & 0.003998 & 1.346049 & 0.006629\\
16 & 43782 & 43688 & 0.560880 & 0.001200 & 0.763372 & 0.004110 & 1.361024 & 0.006883\\
17 & 41494 & 41252 & 0.554893 & 0.001191 & 0.759766 & 0.004222 & 1.369211 & 0.007157\\
18 & 39251 & 39223 & 0.549478 & 0.001180 & 0.749688 & 0.005022 & 1.364364 & 0.008701\\
19 & 37319 & 37254 & 0.544594 & 0.001176 & 0.750020 & 0.004910 & 1.377211 & 0.008573\\
20 & 35271 & 35461 & 0.540129 & 0.001163 & 0.742534 & 0.004394 & 1.374735 & 0.007726\\
21 & 33631 & 33911 & 0.534861 & 0.001160 & 0.744475 & 0.004487 & 1.391904 & 0.007971\\
22 & 31919 & 32033 & 0.527856 & 0.001147 & 0.730673 & 0.004542 & 1.384227 & 0.008201\\
23 & 29574 & 30082 & 0.525044 & 0.001156 & 0.724542 & 0.004544 & 1.379965 & 0.008286\\
24 & 28327 & 28027 & 0.523920 & 0.001164 & 0.739250 & 0.004778 & 1.410999 & 0.008746\\
25 & 26261 & 26270 & 0.519103 & 0.001178 & 0.731693 & 0.004941 & 1.409533 & 0.009100\\
26 & 24236 & 24286 & 0.515637 & 0.001194 & 0.721480 & 0.005048 & 1.399202 & 0.009372\\
27 & 22216 & 22149 & 0.513086 & 0.001210 & 0.712708 & 0.005038 & 1.389063 & 0.009505\\
28 & 20090 & 20242 & 0.508154 & 0.001245 & 0.709395 & 0.005423 & 1.396025 & 0.010251\\
29 & 17962 & 17742 & 0.506019 & 0.001289 & 0.715578 & 0.005672 & 1.414131 & 0.010840\\
30 & 15847 & 15830 & 0.503730 & 0.001335 & 0.706387 & 0.006061 & 1.402312 & 0.011596\\
31 & 13799 & 13596 & 0.500875 & 0.001407 & 0.700708 & 0.006403 & 1.398968 & 0.012350\\
32 & 11735 & 11769 & 0.498915 & 0.001502 & 0.708867 & 0.006986 & 1.420816 & 0.013560\\
33 & 9800 & 9755 & 0.493398 & 0.001597 & 0.699422 & 0.007500 & 1.417561 & 0.014749\\
34 & 8329 & 8361 & 0.490842 & 0.001706 & 0.707928 & 0.008131 & 1.442271 & 0.016099\\
35 & 6828 & 6813 & 0.489287 & 0.001848 & 0.696945 & 0.008724 & 1.424408 & 0.017463\\
36 & 5410 & 5472 & 0.485029 & 0.002012 & 0.700493 & 0.010306 & 1.444228 & 0.020622\\
37 & 4306 & 4419 & 0.482083 & 0.002231 & 0.681407 & 0.010698 & 1.413464 & 0.021630\\
38 & 3439 & 3362 & 0.481093 & 0.002500 & 0.697546 & 0.012524 & 1.449918 & 0.025384\\
39 & 2641 & 2684 & 0.476719 & 0.002739 & 0.673639 & 0.013763 & 1.413074 & 0.027993\\
40 & 1880 & 2004 & 0.481108 & 0.003222 & 0.662584 & 0.015716 & 1.377204 & 0.031513\\
41 & 1431 & 1466 & 0.482852 & 0.003551 & 0.653088 & 0.017247 & 1.352564 & 0.035127\\
42 & 1031 & 1056 & 0.471757 & 0.004321 & 0.655124 & 0.020566 & 1.388689 & 0.042379\\
43 & 795 & 826 & 0.467159 & 0.004673 & 0.698727 & 0.027814 & 1.495695 & 0.057086\\
44 & 546 & 570 & 0.467466 & 0.005603 & 0.655966 & 0.030044 & 1.403238 & 0.061936\\
\bottomrule\end{longtable}\endgroup
\subsection*{Tune: Monash}
\begingroup\scriptsize\setlength{\tabcolsep}{3pt}
\begin{longtable}{rrrrrrrrr}
\toprule $m$ & $N_+$ & $N_-$ & $R_\mu$ & SE & $R_\sigma$ & SE & $R_C$ & SE\\\midrule\endhead
3 & 275233 & 275713 & 0.846884 & 0.001180 & 0.930055 & 0.012077 & 1.098209 & 0.014042\\
4 & 206553 & 207062 & 0.804105 & 0.001157 & 0.920656 & 0.004890 & 1.144946 & 0.005840\\
5 & 167209 & 167031 & 0.769164 & 0.001128 & 0.908432 & 0.002920 & 1.181064 & 0.003485\\
6 & 140141 & 139716 & 0.739063 & 0.001103 & 0.899588 & 0.002882 & 1.217200 & 0.003591\\
7 & 120188 & 120237 & 0.717501 & 0.001084 & 0.894395 & 0.003071 & 1.246541 & 0.003963\\
8 & 105786 & 105743 & 0.698989 & 0.001065 & 0.879145 & 0.003189 & 1.257739 & 0.004253\\
9 & 93406 & 93594 & 0.680393 & 0.001046 & 0.871096 & 0.003322 & 1.280284 & 0.004575\\
10 & 83218 & 82401 & 0.669671 & 0.001044 & 0.872583 & 0.003419 & 1.303003 & 0.004812\\
11 & 73376 & 73523 & 0.657639 & 0.001040 & 0.865593 & 0.003731 & 1.316213 & 0.005366\\
12 & 64322 & 64069 & 0.642013 & 0.001049 & 0.860347 & 0.003772 & 1.340077 & 0.005575\\
13 & 55759 & 55159 & 0.632538 & 0.001069 & 0.844094 & 0.004134 & 1.334455 & 0.006206\\
14 & 45849 & 45900 & 0.624431 & 0.001122 & 0.842627 & 0.004638 & 1.349432 & 0.007098\\
15 & 37436 & 37199 & 0.615408 & 0.001184 & 0.831998 & 0.004693 & 1.351945 & 0.007285\\
16 & 29539 & 29363 & 0.606323 & 0.001273 & 0.830834 & 0.005290 & 1.370282 & 0.008333\\
17 & 22488 & 22257 & 0.599360 & 0.001404 & 0.821512 & 0.005912 & 1.370649 & 0.009486\\
18 & 16298 & 16398 & 0.593253 & 0.001603 & 0.821929 & 0.006983 & 1.385462 & 0.011319\\
19 & 11523 & 11677 & 0.583401 & 0.001803 & 0.817585 & 0.008226 & 1.401412 & 0.013579\\
20 & 7743 & 7693 & 0.572896 & 0.002115 & 0.791503 & 0.009455 & 1.381582 & 0.015937\\
21 & 5189 & 5017 & 0.566924 & 0.002573 & 0.780597 & 0.018717 & 1.376898 & 0.031946\\
22 & 3172 & 3182 & 0.559319 & 0.003081 & 0.801489 & 0.015054 & 1.432973 & 0.026086\\
23 & 1970 & 1909 & 0.563993 & 0.003939 & 0.822246 & 0.019571 & 1.457899 & 0.033906\\
24 & 1104 & 1268 & 0.554724 & 0.004754 & 0.803916 & 0.025371 & 1.449219 & 0.044238\\
25 & 633 & 652 & 0.547210 & 0.006238 & 0.762843 & 0.031252 & 1.394059 & 0.055158\\
\bottomrule\end{longtable}\endgroup

\section{Declaration on historical production records and conditional new production}
For the historical baseline samples, production-time records of the actual
random-seed values and effective lifetime/decay settings, including
\texttt{ParticleDecays:limitTau0} and \texttt{ParticleDecays:tau0Max}, were not
retained. Candidate generation sources survive, but their correspondence to
each historical run cannot be certified. Complete reproduction of the original
event-generation chain, including exact replay of the historical samples,
therefore cannot currently be achieved from the retained records. Here the
lifetime setting concerns the decay prescription, not specifically tau-lepton
decays. Reanalysis of the retained data and reproduction of the supplied
conditional numerical results remain possible; neither establishes the missing
historical generator settings. The separate, stronger records for corrected
CP5 and the remaining Monash/A14 top-up linkage gap are documented above.

If the editor or referees consider new production essential for assessment,
we are prepared to generate fresh samples with explicitly documented physics
and decay settings, independent recorded seeds, complete settings and particle
dumps, and production logs, and to repeat the affected analyses and update the
manuscript accordingly. Such production would provide new documented samples
rather than recover the unknown history of the existing samples. Full
replacement production is deferred and has not been started.

\section{Effect of normalization on conditional widths}
\label{sec:s1_normalization}
For CR state $s$, define $G_s=\langle N_{\rm ch}\rangle_s$ over the complete INEL$>0$-like selected sample, including $N_{\rm ch}=0$ and all MPI counts. At fixed integer $m$, $C_{m,s}=\sigma_{m,s}/\mu_{m,s}$ and $W_{m,s}=\sigma_{m,s}/G_s$. Consequently $R_{C,m}=R_{\sigma,m}/R_{\mu,m}$ and $R_{W,m}=R_{\sigma,m}/R_G$, with $R_G=G_{\rm ON}/G_{\rm OFF}$. A common fixed normalization preserves width ordering; separate CR-state means can change it. The self-normalized MPI coordinates are also supplied, but integer $m$, rather than the normalized coordinate, is matched here.
\begin{center}\small\begin{tabular}{llrrrrr}\toprule Kind & Sample & $B$ & $R_G$ & $R_\sigma<1$ & $R_C>1$ & $R_W>1$\\\midrule
baseline & 13 TeV & 23 & 0.724866 & 23 & 23 & 23\\
baseline & 2.76 TeV & 16 & 0.791969 & 16 & 16 & 14\\
baseline & 7 TeV & 20 & 0.749935 & 20 & 20 & 20\\
tune & ATLAS A14 & 24 & 0.707613 & 24 & 24 & 24\\
tune & CMS CP5 & 42 & 0.615997 & 42 & 42 & 42\\
tune & Monash & 23 & 0.728878 & 23 & 23 & 23\\
\bottomrule\end{tabular}\end{center}
Here $B$ counts retained matched bins. In the 13 TeV baseline and all three independent tune comparisons, absolute narrowing and inclusive-relative broadening coexist in every retained bin. Across the six comparison groups, inclusive-relative broadening holds centrally in 146 of 148 bins, whereas local-relative broadening holds in all 148. The associated marginal 95\% lower bounds for $R_W$ exceed one in 137 bins; this is not a simultaneous confidence statement.
For a conditional standard deviation and an inclusive mean estimated in the same sample, their covariance must be retained. Let $p_m=P(M=m)$, $v_m=\sigma_m^2$, and $X=N_{\rm ch}$. The empirical influence functions used are
\begin{align}
\psi_{\sigma_m}(X,M)&=\frac{\mathbf{1}_{M=m}}{p_m}\frac{(X-\mu_m)^2-v_m}{2\sigma_m},\\
\psi_{W_m}(X,M)&=\frac{\psi_{\sigma_m}(X,M)}{G}-\frac{\sigma_m}{G^2}(X-G).
\end{align}
The estimated variance of $\widehat W_m$ is the empirical mean of $\psi_{W_m}^2$ divided by the total selected sample size. ON/OFF ratio errors assume independent CR samples; historical seed gaps limit independent certification of that assumption. The influence function passed a numerical perturbation check. All normalization ratio identities were checked in all 148 matched bins.
Weber et al., arXiv:1811.07744v2, Sections 2--3 and Fig.~2, already discuss conditional fluctuations alongside self-normalized relations. Their conditional-width discussion does not specify a numerical standard-deviation estimator or supply a conditional-width table with uncertainties. The present test establishes normalization-based compatibility on our retained events, not an exact matched-configuration reproduction or an unambiguous identification of their width metric. Their calculation uses PYTHIA 8.230, non-diffractive collisions, and ALICE-defined primary charged particles at $|\eta|<1$. Our selections and historical primary-particle limitations remain distinct. The added empirical information is separate mean and absolute-width suppression with uncertainty, not the algebraic ratio identities or discovery of conditional fluctuations.
Run \path{python conditional_width_analysis/code/compare_normalizations.py} from the S2 root, or use \path{bash RUN_ANALYSIS.sh}. The inputs are the complete integer joint counts and the reproduced central matched-bin table. The following tables give $R_W$ and one-standard-error uncertainties for every retained matched bin; \path{conditional_width_analysis/results/matched_normalization_comparison.csv} additionally gives $R_C$, $R_\sigma$ and marginal confidence bounds.
\subsection*{Baseline: 13 TeV}
\begin{longtable}{rrrrrrr}\toprule $m$ & $R_\sigma$ & SE & $R_C$ & SE & $R_W$ & SE\\\midrule\endhead
3 & 0.942415 & 0.002311 & 1.113596 & 0.002428 & 1.300123 & 0.003297\\
4 & 0.924434 & 0.002651 & 1.153963 & 0.003006 & 1.275318 & 0.003746\\
5 & 0.900605 & 0.006515 & 1.175887 & 0.008271 & 1.242443 & 0.009015\\
6 & 0.899477 & 0.002877 & 1.221130 & 0.003604 & 1.240888 & 0.004046\\
7 & 0.885776 & 0.003099 & 1.242191 & 0.004038 & 1.221986 & 0.004343\\
8 & 0.878552 & 0.003208 & 1.268541 & 0.004333 & 1.212020 & 0.004490\\
9 & 0.869323 & 0.003271 & 1.280010 & 0.004516 & 1.199288 & 0.004574\\
10 & 0.862401 & 0.003371 & 1.300100 & 0.004795 & 1.189738 & 0.004710\\
11 & 0.855042 & 0.003557 & 1.316127 & 0.005172 & 1.179586 & 0.004962\\
12 & 0.847072 & 0.003745 & 1.327580 & 0.005580 & 1.168591 & 0.005218\\
13 & 0.848811 & 0.003951 & 1.348433 & 0.005991 & 1.170990 & 0.005501\\
14 & 0.839172 & 0.004285 & 1.356427 & 0.006602 & 1.157692 & 0.005955\\
15 & 0.838126 & 0.004782 & 1.371222 & 0.007479 & 1.156249 & 0.006636\\
16 & 0.822711 & 0.005340 & 1.366753 & 0.008525 & 1.134983 & 0.007401\\
17 & 0.821777 & 0.005994 & 1.379790 & 0.009663 & 1.133695 & 0.008299\\
18 & 0.809200 & 0.006739 & 1.379038 & 0.011070 & 1.116344 & 0.009323\\
19 & 0.814684 & 0.008167 & 1.400748 & 0.013478 & 1.123909 & 0.011288\\
20 & 0.799838 & 0.009699 & 1.402663 & 0.016354 & 1.103428 & 0.013398\\
21 & 0.801273 & 0.012007 & 1.420088 & 0.020594 & 1.105409 & 0.016579\\
22 & 0.783353 & 0.014519 & 1.400698 & 0.025011 & 1.080686 & 0.020041\\
23 & 0.798615 & 0.018541 & 1.435806 & 0.032189 & 1.101741 & 0.025588\\
24 & 0.789847 & 0.024390 & 1.435668 & 0.043155 & 1.089645 & 0.033655\\
25 & 0.751916 & 0.032176 & 1.356705 & 0.056256 & 1.037318 & 0.044393\\
\bottomrule\end{longtable}
\subsection*{Baseline: 2.76 TeV}
\begin{longtable}{rrrrrrr}\toprule $m$ & $R_\sigma$ & SE & $R_C$ & SE & $R_W$ & SE\\\midrule\endhead
3 & 0.935319 & 0.002190 & 1.094591 & 0.002335 & 1.181005 & 0.002834\\
4 & 0.911399 & 0.002364 & 1.126993 & 0.002704 & 1.150802 & 0.003044\\
5 & 0.900855 & 0.002449 & 1.166223 & 0.002938 & 1.137488 & 0.003146\\
6 & 0.884945 & 0.002616 & 1.191681 & 0.003294 & 1.117399 & 0.003353\\
7 & 0.874083 & 0.002897 & 1.217928 & 0.003791 & 1.103684 & 0.003700\\
8 & 0.864506 & 0.003130 & 1.240077 & 0.004258 & 1.091590 & 0.003992\\
9 & 0.853173 & 0.003972 & 1.258410 & 0.005594 & 1.077282 & 0.005043\\
10 & 0.849730 & 0.004197 & 1.280682 & 0.005989 & 1.072933 & 0.005326\\
11 & 0.836897 & 0.004852 & 1.292380 & 0.007154 & 1.056730 & 0.006150\\
12 & 0.836750 & 0.006118 & 1.308489 & 0.009152 & 1.056544 & 0.007743\\
13 & 0.833400 & 0.007555 & 1.318909 & 0.011477 & 1.052314 & 0.009554\\
14 & 0.812526 & 0.009517 & 1.308008 & 0.014841 & 1.025957 & 0.012029\\
15 & 0.811625 & 0.012951 & 1.325269 & 0.020390 & 1.024820 & 0.016361\\
16 & 0.790565 & 0.016593 & 1.321873 & 0.026778 & 0.998228 & 0.020957\\
17 & 0.806793 & 0.022864 & 1.359958 & 0.038136 & 1.018719 & 0.028876\\
18 & 0.757660 & 0.034694 & 1.298231 & 0.057917 & 0.956679 & 0.043809\\
\bottomrule\end{longtable}
\subsection*{Baseline: 7 TeV}
\begin{longtable}{rrrrrrr}\toprule $m$ & $R_\sigma$ & SE & $R_C$ & SE & $R_W$ & SE\\\midrule\endhead
3 & 0.936243 & 0.002326 & 1.104508 & 0.002471 & 1.248431 & 0.003190\\
4 & 0.918473 & 0.002443 & 1.142474 & 0.002765 & 1.224736 & 0.003338\\
5 & 0.903264 & 0.002620 & 1.180809 & 0.003148 & 1.204455 & 0.003565\\
6 & 0.890752 & 0.002834 & 1.207075 & 0.003567 & 1.187772 & 0.003842\\
7 & 0.879869 & 0.003406 & 1.233585 & 0.004520 & 1.173260 & 0.004592\\
8 & 0.878612 & 0.003055 & 1.265952 & 0.004138 & 1.171583 & 0.004130\\
9 & 0.866547 & 0.003246 & 1.279657 & 0.004520 & 1.155495 & 0.004380\\
10 & 0.860437 & 0.003428 & 1.299019 & 0.004900 & 1.147347 & 0.004620\\
11 & 0.849836 & 0.003929 & 1.306883 & 0.005757 & 1.133212 & 0.005280\\
12 & 0.843531 & 0.004058 & 1.325215 & 0.006094 & 1.124805 & 0.005451\\
13 & 0.840879 & 0.004600 & 1.337498 & 0.006992 & 1.121268 & 0.006169\\
14 & 0.816053 & 0.005210 & 1.329398 & 0.008154 & 1.088164 & 0.006976\\
15 & 0.828411 & 0.006176 & 1.360332 & 0.009722 & 1.104643 & 0.008260\\
16 & 0.813792 & 0.007469 & 1.364342 & 0.012009 & 1.085149 & 0.009979\\
17 & 0.797132 & 0.008759 & 1.350445 & 0.014387 & 1.062934 & 0.011697\\
18 & 0.801313 & 0.011588 & 1.375013 & 0.019153 & 1.068509 & 0.015464\\
19 & 0.811940 & 0.014606 & 1.408429 & 0.024410 & 1.082680 & 0.019487\\
20 & 0.786016 & 0.018264 & 1.370443 & 0.031419 & 1.048112 & 0.024363\\
21 & 0.774401 & 0.025937 & 1.391036 & 0.044277 & 1.032624 & 0.034589\\
22 & 0.784353 & 0.033715 & 1.394395 & 0.058748 & 1.045894 & 0.044962\\
\bottomrule\end{longtable}
\subsection*{Tune: ATLAS A14}
\begin{longtable}{rrrrrrr}\toprule $m$ & $R_\sigma$ & SE & $R_C$ & SE & $R_W$ & SE\\\midrule\endhead
3 & 0.938461 & 0.003239 & 1.114188 & 0.003591 & 1.326235 & 0.004661\\
4 & 0.919167 & 0.002535 & 1.149773 & 0.002890 & 1.298968 & 0.003686\\
5 & 0.902680 & 0.004318 & 1.182511 & 0.005390 & 1.275669 & 0.006156\\
6 & 0.895926 & 0.002883 & 1.227398 & 0.003664 & 1.266124 & 0.004160\\
7 & 0.886332 & 0.003000 & 1.254090 & 0.003964 & 1.252565 & 0.004319\\
8 & 0.871143 & 0.003543 & 1.270493 & 0.004893 & 1.231101 & 0.005070\\
9 & 0.869066 & 0.005643 & 1.298835 & 0.008139 & 1.228165 & 0.008009\\
10 & 0.864631 & 0.004417 & 1.318940 & 0.006473 & 1.221898 & 0.006290\\
11 & 0.869580 & 0.003590 & 1.354071 & 0.005303 & 1.228891 & 0.005136\\
12 & 0.862624 & 0.003681 & 1.362151 & 0.005538 & 1.219062 & 0.005263\\
13 & 0.854297 & 0.003918 & 1.374653 & 0.006006 & 1.207295 & 0.005592\\
14 & 0.853816 & 0.004148 & 1.393165 & 0.006467 & 1.206614 & 0.005914\\
15 & 0.830182 & 0.005042 & 1.375391 & 0.008049 & 1.173215 & 0.007164\\
16 & 0.830264 & 0.005468 & 1.395118 & 0.008870 & 1.173330 & 0.007763\\
17 & 0.826986 & 0.005265 & 1.402194 & 0.008580 & 1.168698 & 0.007480\\
18 & 0.830861 & 0.006031 & 1.432065 & 0.010004 & 1.174173 & 0.008557\\
19 & 0.811460 & 0.006756 & 1.407576 & 0.011359 & 1.146756 & 0.009578\\
20 & 0.822450 & 0.008100 & 1.446903 & 0.013783 & 1.162288 & 0.011472\\
21 & 0.838581 & 0.009972 & 1.497169 & 0.017199 & 1.185084 & 0.014114\\
22 & 0.797062 & 0.011477 & 1.439904 & 0.020014 & 1.126409 & 0.016235\\
23 & 0.812183 & 0.014428 & 1.468398 & 0.025375 & 1.147779 & 0.020403\\
24 & 0.797101 & 0.016870 & 1.474396 & 0.030410 & 1.126465 & 0.023853\\
25 & 0.831668 & 0.022768 & 1.518746 & 0.040871 & 1.175314 & 0.032186\\
26 & 0.723553 & 0.026867 & 1.379745 & 0.049986 & 1.022526 & 0.037974\\
\bottomrule\end{longtable}
\subsection*{Tune: CMS CP5}
\begin{longtable}{rrrrrrr}\toprule $m$ & $R_\sigma$ & SE & $R_C$ & SE & $R_W$ & SE\\\midrule\endhead
3 & 0.892877 & 0.002761 & 1.083098 & 0.002926 & 1.449483 & 0.004614\\
4 & 0.868585 & 0.002804 & 1.118384 & 0.003177 & 1.410049 & 0.004673\\
5 & 0.847436 & 0.002999 & 1.152730 & 0.003634 & 1.375715 & 0.004976\\
6 & 0.832992 & 0.003179 & 1.191783 & 0.004089 & 1.352267 & 0.005258\\
7 & 0.818282 & 0.003347 & 1.215121 & 0.004525 & 1.328388 & 0.005522\\
8 & 0.806547 & 0.003374 & 1.237857 & 0.004732 & 1.309337 & 0.005563\\
9 & 0.799898 & 0.003620 & 1.260554 & 0.005214 & 1.298544 & 0.005952\\
10 & 0.785732 & 0.004243 & 1.272887 & 0.006443 & 1.275546 & 0.006951\\
11 & 0.780998 & 0.003863 & 1.290572 & 0.005932 & 1.267861 & 0.006339\\
12 & 0.775451 & 0.003811 & 1.308579 & 0.005981 & 1.258856 & 0.006255\\
13 & 0.763939 & 0.011121 & 1.310397 & 0.018629 & 1.240167 & 0.018069\\
14 & 0.767835 & 0.004434 & 1.335887 & 0.007272 & 1.246492 & 0.007254\\
15 & 0.762605 & 0.003998 & 1.346049 & 0.006629 & 1.238002 & 0.006554\\
16 & 0.763372 & 0.004110 & 1.361024 & 0.006883 & 1.239247 & 0.006733\\
17 & 0.759766 & 0.004222 & 1.369211 & 0.007157 & 1.233393 & 0.006911\\
18 & 0.749688 & 0.005022 & 1.364364 & 0.008701 & 1.217033 & 0.008198\\
19 & 0.750020 & 0.004910 & 1.377211 & 0.008573 & 1.217572 & 0.008018\\
20 & 0.742534 & 0.004394 & 1.374735 & 0.007726 & 1.205420 & 0.007186\\
21 & 0.744475 & 0.004487 & 1.391904 & 0.007971 & 1.208570 & 0.007337\\
22 & 0.730673 & 0.004542 & 1.384227 & 0.008201 & 1.186165 & 0.007424\\
23 & 0.724542 & 0.004544 & 1.379965 & 0.008286 & 1.176210 & 0.007427\\
24 & 0.739250 & 0.004778 & 1.410999 & 0.008746 & 1.200088 & 0.007806\\
25 & 0.731693 & 0.004941 & 1.409533 & 0.009100 & 1.187819 & 0.008067\\
26 & 0.721480 & 0.005048 & 1.399202 & 0.009372 & 1.171241 & 0.008239\\
27 & 0.712708 & 0.005038 & 1.389063 & 0.009505 & 1.157001 & 0.008223\\
28 & 0.709395 & 0.005423 & 1.396025 & 0.010251 & 1.151622 & 0.008843\\
29 & 0.715578 & 0.005672 & 1.414131 & 0.010840 & 1.161658 & 0.009249\\
30 & 0.706387 & 0.006061 & 1.402312 & 0.011596 & 1.146738 & 0.009875\\
31 & 0.700708 & 0.006403 & 1.398968 & 0.012350 & 1.137520 & 0.010428\\
32 & 0.708867 & 0.006986 & 1.420816 & 0.013560 & 1.150764 & 0.011372\\
33 & 0.699422 & 0.007500 & 1.417561 & 0.014749 & 1.135431 & 0.012204\\
34 & 0.707928 & 0.008131 & 1.442271 & 0.016099 & 1.149240 & 0.013227\\
35 & 0.696945 & 0.008724 & 1.424408 & 0.017463 & 1.131410 & 0.014189\\
36 & 0.700493 & 0.010306 & 1.444228 & 0.020622 & 1.137170 & 0.016752\\
37 & 0.681407 & 0.010698 & 1.413464 & 0.021630 & 1.106186 & 0.017386\\
38 & 0.697546 & 0.012524 & 1.449918 & 0.025384 & 1.132385 & 0.020349\\
39 & 0.673639 & 0.013763 & 1.413074 & 0.027993 & 1.093575 & 0.022358\\
40 & 0.662584 & 0.015716 & 1.377204 & 0.031513 & 1.075629 & 0.025525\\
41 & 0.653088 & 0.017247 & 1.352564 & 0.035127 & 1.060214 & 0.028011\\
42 & 0.655124 & 0.020566 & 1.388689 & 0.042379 & 1.063518 & 0.033396\\
43 & 0.698727 & 0.027814 & 1.495695 & 0.057086 & 1.134303 & 0.045160\\
44 & 0.655966 & 0.030044 & 1.403238 & 0.061936 & 1.064886 & 0.048778\\
\bottomrule\end{longtable}
\subsection*{Tune: Monash}
\begin{longtable}{rrrrrrr}\toprule $m$ & $R_\sigma$ & SE & $R_C$ & SE & $R_W$ & SE\\\midrule\endhead
3 & 0.930055 & 0.012077 & 1.098209 & 0.014042 & 1.276010 & 0.016579\\
4 & 0.920656 & 0.004890 & 1.144946 & 0.005840 & 1.263115 & 0.006753\\
5 & 0.908432 & 0.002920 & 1.181064 & 0.003485 & 1.246343 & 0.004083\\
6 & 0.899588 & 0.002882 & 1.217200 & 0.003591 & 1.234209 & 0.004030\\
7 & 0.894395 & 0.003071 & 1.246541 & 0.003963 & 1.227084 & 0.004283\\
8 & 0.879145 & 0.003189 & 1.257739 & 0.004253 & 1.206163 & 0.004440\\
9 & 0.871096 & 0.003322 & 1.280284 & 0.004575 & 1.195119 & 0.004618\\
10 & 0.872583 & 0.003419 & 1.303003 & 0.004812 & 1.197160 & 0.004751\\
11 & 0.865593 & 0.003731 & 1.316213 & 0.005366 & 1.187569 & 0.005172\\
12 & 0.860347 & 0.003772 & 1.340077 & 0.005575 & 1.180372 & 0.005228\\
13 & 0.844094 & 0.004134 & 1.334455 & 0.006206 & 1.158074 & 0.005716\\
14 & 0.842627 & 0.004638 & 1.349432 & 0.007098 & 1.156061 & 0.006402\\
15 & 0.831998 & 0.004693 & 1.351945 & 0.007285 & 1.141478 & 0.006478\\
16 & 0.830834 & 0.005290 & 1.370282 & 0.008333 & 1.139880 & 0.007292\\
17 & 0.821512 & 0.005912 & 1.370649 & 0.009486 & 1.127092 & 0.008142\\
18 & 0.821929 & 0.006983 & 1.385462 & 0.011319 & 1.127663 & 0.009607\\
19 & 0.817585 & 0.008226 & 1.401412 & 0.013579 & 1.121704 & 0.011307\\
20 & 0.791503 & 0.009455 & 1.381582 & 0.015937 & 1.085920 & 0.012990\\
21 & 0.780597 & 0.018717 & 1.376898 & 0.031946 & 1.070957 & 0.025685\\
22 & 0.801489 & 0.015054 & 1.432973 & 0.026086 & 1.099621 & 0.020665\\
23 & 0.822246 & 0.019571 & 1.457899 & 0.033906 & 1.128098 & 0.026860\\
24 & 0.803916 & 0.025371 & 1.449219 & 0.044238 & 1.102950 & 0.034815\\
25 & 0.762843 & 0.031252 & 1.394059 & 0.055158 & 1.046599 & 0.042881\\
\bottomrule\end{longtable}

\section{AI assistance and validation record}
The manuscript Methods subsection on AI-assisted code development and validation identifies OpenAI Codex, the assisted programming, diagnostic, plotting and editorial tasks, and the executable validation performed. Selected retained project-related metadata record \texttt{gpt-5.5}, \texttt{gpt-5.6-sol}, \texttt{gpt-5.6-luna}, \texttt{gpt-6-astra} and \texttt{gpt-6.1-sol}; the current revision session records \texttt{gpt-6.1-sol} through the Codex VS Code integration, client \texttt{0.162.0-alpha.17.2}. These are recorded identifiers, not independently verified immutable snapshots or a complete model-to-artifact history.
S2 supplies \path{ai_disclosure/LOCAL_MODEL_RECORDS.json} and \path{ai_disclosure/TASK_VALIDATION_MAP.csv}. Private conversation bodies are not included. The available numerical evidence includes the 60-million-event central reanalysis, full six-baseline particle scan, fourth-moment and empirical-bootstrap errors, normalization influence-function perturbation check, source checks, validation-only generator tests, and fresh-archive numerical/plot reproduction. Automated checks were themselves developed with AI assistance and are not independent human validation. They verify numerical consistency with retained inputs and do not restore missing historical generation records. Scientific interpretation and responsibility remain with the authors.